\documentclass[12pt]{article}
\usepackage{epsfig}

\newcommand{\bce}{\begin{center}}
\newcommand{\ece}{\end{center}}
\newcommand{\be}{\begin{equation}}
\newcommand{\ee}{\end{equation}}
\newcommand{\bea}{\begin{eqnarray}}

\newcommand{\eea}{\end{eqnarray}}

\newcommand{\ba}{\begin{array}}
\newcommand{\ea}{\end{array}}
\newcommand{\bnabla}{\mbox{\boldmath $\nabla$}}

\newcommand{\bDelta}{\mbox{\boldmath $\Delta$}}

\newcommand{\bPhi}{\mbox{\boldmath $\Phi$}}

\newcommand{\bkappa}{\mbox{\boldmath $\kappa$}}
\newcommand{\bfkappa}{\mbox{\boldmath $\kappa$}}

\newcommand{\bfe}{\mbox{\boldmath $e$}}

\newcommand{\bfb}{\mbox{\boldmath $b$}}
\newcommand{\br}{\mbox{\boldmath $r$}}
\newcommand{\bfr}{\mbox{\boldmath $r$}}
\newcommand{\bk}{\mbox{\boldmath $k$}}
\newcommand{\bfk}{\mbox{\boldmath $k$}}

\newcommand{\singlespace}{
    \renewcommand{\baselinestretch}{1.1}\large\normalsize}

\def\lsim{\mathrel{\rlap{\lower4pt\hbox{\hskip1pt$\sim$}}
    \raise1pt\hbox{$<$}}}         %less than or approx. symbol
\def\gsim{\mathrel{\rlap{\lower4pt\hbox{\hskip1pt$\sim$}}
    \raise1pt\hbox{$>$}}}         %greater than or approx. symbol

\def\Pom{{\bf I\!P}}

\newcommand{\dst}{\displaystyle}

\def\Pom{{\bf I\!P}}

\def\lsim{\mathrel{\rlap{\lower4pt\hbox{\hskip1pt$\sim$}}
    \raise1pt\hbox{$<$}}}         %less than or approx. symbol

\def\gsim{\mathrel{\rlap{\lower4pt\hbox{\hskip1pt$\sim$}}
    \raise1pt\hbox{$>$}}}         %greater than or approx. symbol

\date{}
\begin{document}
\begin{flushright}
{FZJ-IKP(TH)-2000-23}
\end{flushright}
\vspace{1.0cm}

\begin{center}

{\Large \bf  Coherent
production of hard dijets on nuclei in QCD
\vspace{1.0cm}}\\

{\large %\bf 
N.N. Nikolaev$^{1,2}$, W. Sch\"afer$^{1}$ and  G. Schwiete$^{1}$}\\

\vspace{0.5cm} $^{1}${ \em Institut  f\"ur
Kernphysik (Theorie), Forschungszentrum J\"ulich,\\ D-52425 J\"ulich, Germany\\}

$^{2}${ \em L.D.Landau Institute for Theoretical Physics\\ Chernogolovka,
Moscow Region 142 432, Russia}\\

\vspace{0.5cm}

\end{center}

%\begin{document}
\large
%\maketitle
%\vspace{-9cm}
%\makebox[\textwidth][r]{\large\bf FZJ-IKP(Th)-2000/08}
%\vspace{8cm}

\begin{abstract}
We formulate the perturbative QCD approach to coherent diffractive dijet
production in pion-nucleon and pion-nucleus collisions at high energy.
For hard dijets the Pomeron splitting mechanism in which both 
helicity amplitudes are proportional to the unintegrated 
gluon structure function of the proton  ${\cal F}(x,\bk^2)$ and pion 
distribution amplitude $\phi_{\pi}(z)$ is shown to dominate. In
nuclear diffraction multiple Pomeron splitting components are found 
to give antishadowing contributions at large jet momentum $\bk$. 
To leading twist there is an exact cancelation of effects of nuclear 
attenuation and antishadowing/broadening of multiple Pomeron splitting
contributions. The next-to-leading higher twist correction driven
by nuclear rescatterings is calculable in hard QCD and proves to
be numerically very large. We argue that large higher twist effects
do not preclude the determination of gross features of $\phi_{\pi}(z)$.
Our results on the atomic mass number and momentum dependence of dijet
cross sections agree well with the preliminary findings from the E791 
experiment.

\end{abstract}
\newpage
%%%%%%%%%%%%%%%%%%%%%%%%
%%\doublespace
\singlespace
%%%%%%%%%%%%%%%%%%%%%%%%
\section{Introduction}

Ever since the classic work in early 50's
by Landau, Pomeranchuk, Feinberg and
Glauber  on diffraction excitation of deuterons  
\cite{Landau,Feinberg,Glauber} into the proton-neutron 
continuum the momentum spectrum of excitation products (protons \& 
neutrons) is known to be given by the momentum distribution of 
constituents in the deuteron. More recent work on diffraction 
dissociation focused on diffractive deep inelastic scattering (DIS).
Here the microscopic QCD description of diffractive scattering by 
exchange of a color-singlet two-gluon tower in the $t$-channel reveals
a sensitivity of the mass spectrum in diffractive excitation of the 
continuum $q\bar{q}$ states to the gluon structure function of the 
target \cite{NZ92}. Furthermore, extending early considerations in
\cite{NZ92} Nikolaev and Zakharov have 
shown in 1994 \cite{NZsplit} that in diffraction excitation of hard dijets 
$\gamma^{*}p \to p' q\bar{q}$ there exist two regimes depending on 
how the large transverse momentum $\bk$ of the jets compares to the
hard scale of DIS, i.e., whether $k \lsim Q$ or
$k\gsim Q$. In the first regime the transverse momentum $\bk$ of 
jets comes from the intrinsic momentum $\bk$ of the quark (antiquark) 
in the $q\bar{q}$ Fock state of the $\gamma^*$ and diffractive amplitudes 
are proportional to the familiar integrated gluon structure 
function (GSF) of the target proton $G(x,\bk^2)$, see also \cite{GNZcharm}. 
In the second regime, $k\gsim Q$, diffractive dijets are a unique 
probe of the differential (unintegrated) gluon structure function 
(DGSF) of the proton 
${\cal F}(x,Q^2)=\partial G(x,Q^2)/\partial\log Q^2$. Specifically, 
in this regime the transverse momentum $\bk$ of jets is provided 
not by the momentum of $q$ and $\bar{q}$ in the virtual photon, 
but by the momentum of gluons in the Pomeron. Correspondingly, this regime 
of diffractive DIS has been dubbed `the splitting of Pomerons into 
dijets' \cite{NZsplit}. In this regime the diffractive dijet production 
amplitude is proportional to ${\cal F}(x,\bk^2)$. Subsequently 
Golec-Biernat, Kwiecinski and Martin \cite{Martin} reformulated the 
formalism \cite{NZsplit} in terms of the off-diagonal (skewed) parton 
distributions. Because the skewed distributions can be approximated 
\cite{Radyushkin} by the diagonal ones at a rescaled $x$, after this 
rescaling the formulas of \cite{NZsplit} are recovered.  

In the present communication we extend the approach \cite{NZsplit} 
to coherent diffraction of pions into dijets on the nucleon and 
nuclear targets.
The principal novelty compared to photo- and electroproduction is
that the pion-quark-antiquark vertex is non-pointlike which makes
splitting of the Pomeron the ever more important mechanism for hard dijets.  
We focus on coherent diffraction production on nuclei which has recently
been measured by the E791 collaboration \cite{SELEX}. 
The principal issues with the hard QCD interpretation of these data are whether
$1.25 \lsim k \lsim 2.5$ GeV is sufficiently hard for the pQCD treatment,
how large are next-to-leading twist corrections, what are nuclear
effects and whether the extraction of the pion distribution amplitude
is possible from the E791 data. The two major
nuclear effects  one has to deal with are nuclear attenuation and 
nuclear broadening of jets. The practical calculation of diffraction 
on nuclear targets involves evaluation of multiple gluon exchanges 
between nucleon and excited $q\bar{q}$ system and we take full advantage 
of the recent determination of the DGSF of the proton \cite{INDGSF}. 
We demonstrate that the broadening of the jet momentum distribution 
comes entirely from the multiple Pomeron splitting diagrams. The large 
$\bk^2$ behavior of ${\cal F}(x,\bk^2)$ found in \cite{INDGSF} is shown 
to entail a remarkable cancelation of the attenuation and broadening 
effects to leading twist. In view of these cancelations the principal 
nuclear effect is a higher twist correction which is perturbatively
calculable and is proportional to $G(x,\bk^2)$. 

The further presentation is organized as follows. In section 2 we 
introduce the principal formalism starting with excitation of diffractive 
dijets on free nucleons and isolate the two helicity components of the 
diffraction cross section. We demonstrate how the dominance of the
Pomeron splitting mechanism into hard dijets and therefore the proportionality of 
diffraction amplitudes to ${\cal F}(x,\bk^2)$ do emerge because 
of the non-pointlike pion-quark-antiquark coupling. For the same reason
diffractive amplitudes are shown to be proportional to the pion 
lightcone distribution amplitude $\phi_{\pi}(z)$ of much discussion
in the recent literature (for the review see \cite{CZ,Kroll,Brodsky}). The
possibility of measuring $\phi_{\pi}(z)$ in diffraction of pions into 
hard dijets has been mentioned in \cite{FMS} but as we show the claim 
in \cite{FMS} that to the leading twist diffractive amplitudes are proportional to 
$G(x,\bk^2)$ is in error. Calculation of multiple gluon exchange in 
diffraction off nuclei to leading and higher twist is described in 
section 3. The novel feature of nuclear diffraction are multiple-Pomeron
splitting processes in which the $\bk$ distribution is broadened by
the gluon momentum coming from different split Pomerons. In the standard 
nuclear multiple-scattering expansion the higher order nuclear 
rescatterings are known to generate nuclear shadowing \cite{Glauber2}. 
We demonstrate 
that after reformulation in terms of multiple Pomeron splitting components 
the nuclear multiple scattering expansion takes a form in which higher 
order Pomeron splitting components give antishadowing contributions, i.e. 
an enhancement of the corresponding impulse approximation term. 
We find an exact cancelation of effects of nuclear attenuation and 
antishadowing/broadening of multiple Pomeron splitting in the leading 
twist $\bk^2$ distributions. The higher twist correction rises with 
the multiplicity of split Pomerons and is shown to be proportional 
to the integrated gluon structure function of the proton $G(x,\bk^2)$. 
It is an antishadowing correction and rises with the nuclear mass 
number. In Section 5 we summarize our main results and present a 
comparison with the preliminary experimental findings from E791.
Our numerical analysis shows that the leading plus next-to-leading 
twist asymptopia sets in only for $k\gsim $ 2--3 GeV, somewhat beyond
the kinematical range of E791 $1.25 \lsim k \lsim 2.5$ GeV. Our
numerical results for the $k$ and atomic mass number dependence 
of the dijet cross section are consistent with the experimental
findings by E791 \cite{SELEX}.

%-------- Section 2
\section{Microscopic QCD mechanism of diffraction into dijets}

We only need a slight adaptation of the formalism developed in 
\cite{NZ92,NZsplit,NPZLT}. Diffraction dissociation of the pion
into the high mass continuum, hard, $q\bar{q}$ dijet final state,
$$
\pi p \to p' q\bar{q}\,,
$$
is described by the four pQCD diagrams of Fig.~1. In this paper 
the dijet cross section is calculated at the parton level. 
The relevant kinematical variables are
shown in Fig. 1, $\bDelta$ is the transverse momentum of the
excited dijet, quark and antiquark jets carry a fraction $z$ 
and $1-z$ of the pion's  momentum and the invariant mass of the
excited pair $M$ is given by
\begin{equation}
M^{2}={{\bk}^{2}+ m_{f}^{2}\over z(1-z)}\, ,
\label{eq:2.1}
\end{equation}
where $m_{f}$ is the quark mass. Such a parton level modeling of final states 
is applicable if the invariant mass $M$ of the diffractive system
is above the prominent resonances which are excited diffractively 
from pions, specifically $A_{1}(1260)$, $\pi(1670)$, $\pi(2100)$ \cite{Daum} 
and $\pi(1300)$, $\pi(1800)$
\cite{Bellini,Zielinski} and, perhaps, still higher radial and
angular excitations of the pion. For instance, the color dipole
model analysis in \cite{CTCEBAF} has shown that diffraction 
excitation of nucleons is exhausted by resonance excitation 
for $M \lsim 3$ GeV. Therefore, the parton level calculation is 
viable at best for jets with $k \gsim 1.5 $ GeV.

The minor technical difference 
from diffractive excitation of the photon studied in 
\cite{NZ92,NZsplit,NPZLT} is the change from the
pointlike $\gamma^{*}q\bar{q}$ vertex $eA_{\mu}\overline{\Psi} 
\gamma_{\mu}\Psi$ to the non-pointlike $\pi q\bar{q}$ vertex
$i\Gamma_{\pi}(M^2)\overline{\Psi}\gamma_{5}\Psi$. In terms of
the quark \& antiquark helicities $\lambda$ the $\pi \bar{q}(\bk) {q}(-\bk)$ vertex 
has the form (for the related discussion see Jaus \cite{Jaus})
\begin{equation}
\overline{\Psi}_{\lambda}(\bk)\gamma_{5}\Psi_{\bar{\lambda}}(-\bk) 
= {\lambda \over \sqrt{z(1-z)}}
[m_{f} \delta_{\lambda -\bar{\lambda}} - \sqrt{2}\bk\cdot \bfe_{-\lambda}
\delta_{\lambda \bar{\lambda}}]\, ,
\label{eq:2.2}
\end{equation}
where $m_{f}$ is the quark mass and 
$\bfe_{\lambda}=-
(\lambda \bfe_{x}+i\bfe_{y})/\sqrt{2}$. In transitions of spin-zero
pions into $q\bar{q}$ states 
with the sum of helicities $\lambda+\bar{\lambda}=\pm 1$ the latter is compensated
by the orbital angular momentum of quark and antiquark. 

In close analogy to the QCD description of diffractive dijet excitation
in DIS, $\gamma^{*}\to q\bar{q}$ developed in \cite{NZ92,NZsplit,NPZLT},
the two helicity transitions in (\ref{eq:2.2}) define the two diffractive
amplitudes $\Phi_{0}(z,\bk,\bDelta)$ and $\bPhi_{1}(z,\bk,\bDelta)$.
The lower blob in diagrams of Fig.~1 is related to the off-forward and 
off-diagonal differential gluon structure function of the target proton 
${\cal{F}}\left(x_{1},x_{2},\bkappa,\bDelta\right)$. In the considered high 
energy limit the two-gluon exchange interaction of $q\bar{q}$ 
states with the target conserves the quark and antiquark helicities 
exactly. This quark helicity conservation simplifies substantially 
the calculation of multiple Pomeron exchanges in diffraction off nuclei. 
One can readily update to the pion beam an analysis of the $\bDelta$ 
dependence of diffractive amplitudes carried out for diffractive DIS in 
\cite{NPZslope}, but for the purposes of diffraction on nuclei we only
need the amplitudes for $\pi N \to N (q\bar{q})$ in the forward
limit $\bDelta=0$ and suppress $\bDelta$ as an argument of diffractive 
amplitudes wherever it is appropriate. The lightcone momentum of the 
gluon is related to the change of the mass of the diffractive system,
$x_{g}= (M_{f}^2 -M_{in}^2)/W^2$. In the considered problem $M_{in}^2=
m_{\pi}^2$ can be neglected, the diffractive mass $M$ is generated
by the first gluon,
\be
x_{1}\approx x_{\Pom} = {M^2 \over W^2}\, ,
\label{eq:2.3}
\ee
whereas the second exchange changes it only a little and $x_{2}\approx 0$. 
The detailed discussion of kinematics is found in \cite{Martin,Radyushkin}
and need not be repeated here, the principal point is that in the   
diffractive limit of $x_{\Pom}\ll 1$ the relevant off-diagonal 
differential gluon structure function of the target proton can be 
approximated  \cite{Radyushkin,Martin} by the conventional DGSF 
taken at $x={1\over 2} (x_{1}+x_{2}) = {1\over 2}x_{\Pom}$,
i.e.,
\be
{\cal{F}}\left(x_{1},x_{2},\bkappa,\bDelta\right)=
{\cal{F}}({1\over 2}x_{\Pom},\bkappa,\bDelta=0)=
{\partial G({1\over 2}x_{\Pom},\bkappa^2)\over \partial \log \bkappa^2}\, .
\label{eq:2.4}
\ee
After this rescaling one recovers 
precisely the expressions of \cite{NZsplit} for the diffractive amplitudes
$\Phi_{2},\bPhi_{1}$.  
The hard scale in  $q\bar{q}$ excitation is set by the large transverse 
momentum of jets, $\bk^2 \gg 1$ GeV$^2$, and 
it is $\alpha_S(\bk^2)$ which enters the 
gluon-quark and gluon-antiquark vertices in the diffractive amplitudes.

We find it convenient to introduce
\be
\sigma_{0}={4\pi \over 3}\int d^{2}\bk 
{{\cal{F}}({1\over 2}x_{\Pom},\bk^2) \over \bk^4}
\, 
\label{eq:2.5}
\ee
and the distribution function  
\be
f^{(1)}(\bk) = {4\pi \over 3\sigma_{0}} 
{{\cal{F}}({1\over 2}x_{\Pom},{\bk^2}) \over \bk^4}\, ,
\label{eq:2.6}
\ee
normalized to unity: $\int d^2\bk f^{(1)}(\bk) =1$. For the
sake of brevity of notations we suppress the dependence on $x_{\Pom}$.

We define the two diffractive amplitudes $\Phi_{0}(z,\bk)$ and 
$\bPhi_{1}(z,\bk)$ as (we use the normalization slightly different 
from that in \cite{NZsplit,NPZLT}):
\begin{eqnarray}
\Phi_0(z,\bk)&=&\alpha_{S}(\bk^2) \sigma_{0}
\int d^2\bkappa m_{f}\left[\psi_{\pi} (z,\bk)-\psi_{\pi} (z,\bk-\bkappa)
\right] f^{(1)}(\bkappa)
\nonumber\\
&=&\alpha_{S}(\bk^2) \sigma_{0}\left[
\int d^2\bkappa m_{f}\psi_{\pi} (z,\bk) f^{(1)}(\bkappa)
- \int d^2\bkappa
m_{f}\psi_{\pi} (z,\bkappa)f^{(1)}(\bk-\bkappa)\right]
\, , \\
\label{eq:2.7}
%\end{eqnarray}
%\begin{eqnarray}
\bPhi_1(z,\bk)&=&\alpha_{S}(\bk^2)\sigma_{0} 
\int d^2\bkappa \left[\bk\psi_{\pi} (z,\bk)-
(\bk-\bkappa)\psi_{\pi} (z,\bk-\bkappa)\right]
 f^{(1)}(\bkappa)
\nonumber\\
&=&\alpha_{S}(\bk^2)\sigma_{0} 
\left[\int d^2\bkappa \bk\psi_{\pi} (z,\bk) f^{(1)}(\bkappa)
- \int d^2\bkappa \bkappa
\psi_{\pi} (z,\bkappa)f^{(1)}(\bk-\bkappa)\right]\, .
\label{eq:2.8}
\end{eqnarray}
The differential cross section of forward dijet production equals
\begin{eqnarray}
\left.{d\sigma_{D} \over dz d\bk^2 d\bDelta^2}\right|_{\bDelta=0}
= {\pi^3 \over 24} \left\{|\Phi_0|^2 + |\bPhi_1|^2 \right\}
\label{eq:2.9}
\end{eqnarray}

The radial wave function of the pion in momentum space is defined
in terms of the $\pi q\bar{q}$ vertex function as
\be
\psi_{\pi}(z,\bk)={N_{c}\Gamma_{\pi}(M^{2}) \over 4\pi^3 z(1-z)( M^2 -m_{\pi}^2)}
\label{eq:2.10}
\ee
and is so normalized that the $\pi \to \mu \nu$
decay constant equals (we use the PDG convention $F_{\pi}=131$ MeV \cite{PDG})
\be
F_{\pi}= \int d^2\bk dz m_{f} \psi_{\pi}(z,\bk)
=F_{\pi}\int dz \phi_{\pi}(z)\, .
\label{eq:2.11}
\ee
Here we indicated also the relationship to the often discussed pion 
distribution amplitude $\phi_{\pi}(z)$ which for the purposes of 
our discussion we find it convenient to normalize to unity, 
$\int dz \phi_{\pi}(z)=1$. We recall that for 
the pointlike photon $\Gamma_{\gamma}(M^{2})=e_{f}$, where $e_{f}$ is
the electric charge of the quark. In contrast to the pointlike 
photon for the non-pointlike pion $\Gamma_{\pi}(M^2)$ vanishes at 
large $M^2$ faster than $\propto M^{-2}$, the relevant arguments
are found in Brodsky \& Lepage \cite{Brodsky} and need not be 
repeated here. To this end we disagree with
Ref.\cite{FMS} in which the pointlike $\Gamma_{\pi}(M^{2})=const$ 
is assigned to the large-$M^2$ tail of the pion wave function.

Let us focus on the amplitude $\Phi_0(z,\bk)$. The first term in 
(\ref{eq:2.7}) comes from diagrams 1a, 1b, the corresponding spectrum of
jets would be identical to the quark (antiquark) momentum distribution in the
pion. Because the second term is a convolution of the gluon 
distribution and wave function, as such it is a broader function 
of $\bk$ than $\psi_{\pi}(z,\bk)$ alone and would always take over at large $\bk$.  
The precise pattern of this dominance depends on the large-$\bk$ 
properties of ${\cal{F}}(x_{\Pom},\bk^2)$, the detailed discussion of which 
is found in \cite{INDGSF}. Here we only mention that for $x\sim 10^{-2}$ relevant
to the E791 experiment the results of \cite{INDGSF} correspond to the
inverse power asymptotics at large-$\bk^2$
\be
f^{(1)}(\bk)\propto k^{-2\delta}
\label{eq:2.12}
\ee
with the exponent $\delta \sim $ 2.15. At $x\sim 10^{-2}$ this asymptotics 
sets in at $\bk^2 \gsim k_h^2 \sim 1$ GeV$^2$,  for smaller values of $x_{\Pom}$ 
the exponent $\delta$ is smaller, for instance $\delta \sim 1.7$
for $x_{\Pom}=10^{-3}$, and $k_h^2$ gets 
larger, see Fig.~2. 
As such, $f^{(1)}(\bk)$ decreases at large $\bk$ much slower than
the pion wave function $\psi_{\pi}(z,\bk)$ (see the explicit parametrisation (\ref{eq:4.2.1}) below)
 and the asymptotics of the convolution will be controlled 
by the asymptotics of $f^{(1)}(\bk)$. We evaluate the second term in (\ref{eq:2.7}) 
to the next-to-leading twist 
making use of the small-$\bkappa$ expansion
\be
f^{(1)}(\bk -\bkappa)\approx   
f^{(1)}(\bk)\left(1+{2\delta \bk\bkappa \over \bk^2}-\delta {\bkappa^2 \over \bk^2}
+2 \delta(\delta+1){(\bk \bkappa)^2 \over \bk^4}\right)
\label{eq:2.13}
\ee
and obtain
\begin{eqnarray}
\Phi_0(z,\bk)= \alpha_{S}(\bk^2)\sigma_{0}\left[
m_{f}\psi_{\pi}(z,\bk) - f^{(1)}(\bk)F_{\pi} \phi_{\pi}(z)
\left(1+\delta^2{\langle \kappa_{\pi}^2(z)\rangle 
\over  \bk^2} \right)\right]\,,
\label{eq:2.14}
\end{eqnarray}
where
\be
\langle \kappa_{\pi}^2(z)\rangle ={\displaystyle \int d^2\bkappa\, \bkappa^2 \psi_{\pi}(z,\bkappa)
\over 
\displaystyle \int d^2\bkappa  \psi_{\pi}(z,\bkappa)}\, .
\label{eq:2.15}
\ee

In the related evaluation of the large-$\bk$ behavior of the convolution 
term in the diffractive amplitude $\bPhi_{1}$ for excitation of dijets with 
the sum of helicities $\lambda+\bar{\lambda}= \pm 1$ the leading term comes
from the second term $\propto \delta 2\bk\bkappa/\bk^2$ in the
expansion (\ref{eq:2.13}). Then we find
\be
\bPhi_{1}(z,\bk)=\alpha_{S}(\bk^2)\sigma_{0}\bk\left[
\psi_{\pi}(z,\bk)-{\delta \langle \kappa_{\pi}^2(z)\rangle  
\over m_{f} \bk^2}f^{(1)}(\bk)F_\pi \phi_{\pi}(z)\right]\, ,
\label{eq:2.16}
\ee
Evidently, in the region of large $\bk$ where $\psi_{\pi}(z,\bk)$ dies out 
the amplitude $\bPhi_{1}$ will give the higher twist correction 
to the high-$\bk$ dijet cross section. 

The resulting large-$\bk$ asymptotics of the differential cross section 
for dijet production on nucleons reads 
\bea
\left.{d\sigma_{D} \over dz d\bk^2 d\bDelta^2}\right|_{\bDelta=0}
= {2\pi^5 \over 27} F^2_{\pi} \phi^2_{\pi}(z)  \alpha^2_S(\bk^2)
\left[{{\cal{F}}({1\over 2}x_{\Pom},\bk) \over \bk^4}\right]^2
\cdot \left\{ 1 + 2 \delta^2{ \langle \kappa_{\pi}^2(z)\rangle \over \bk^2}
\left( 1+ {\langle \kappa_{\pi}^2(z)\rangle \over 2m_{f}^2}\right)\right\}\,.
\label{eq:2.17}
\eea
Evidently, the large-$\bk$ asymptotic behaviour $\propto k^{-8}$ is suggested 
by purely dimensional counting. Substantial departure from the
law $\propto k^{-8}$ is possible because of scaling violations in 
${\cal{F}}({1\over 2}x_{\Pom},\bk)$. According to the recent 
phenomenological analysis \cite{INDGSF}, the DGSF 
${\cal{F}}(x,\bk)$ is approximately constant at moderately small
$x \sim 10^{-2}$, but rises steeply with $\bk^2$ at $x\lsim 10^{-3}$.
Also, the experimental data are taken at fixed $W^2$, so that 
in view of (\ref{eq:2.1}) and (\ref{eq:2.3}) the $\bk^2$ 
dependence of the observed cross section is affected by the 
increase of $x_{\Pom}$ and decrease of ${\cal{F}}({1\over 2}x_{\Pom},
\bk)$ with increasing $\bk^2$. Similar kinematical bias affects the 
$z$-dependence of the experimentally observed cross section.
 
There are three important aspects of our diffractive dijet excitation
amplitudes at large $\bk$ where the pion wave function dies out. 

First, here both helicity amplitudes are proportional to the DGSF 
of the target proton ${\cal{F}}({1\over 2}x_{\Pom},\bk^2)$, i.e. 
the jet momentum comes from the momentum of gluons in the exchanged 
Pomeron, hence the term "splitting the Pomeron". To this end we 
recall that Nikolaev and Zakharov found the same proportionality
of diffractive amplitudes to ${\cal{F}}({1\over 2}x_{\Pom},\bk^2)$
also for real photoproduction with pointlike 
$\gamma q\bar{q}$ QED vertex \cite{NZsplit}. From here one would conclude
that this property does not require the wave function of the 
pion to be soft and  the $\pi q\bar{q}$ vertex function $\Gamma_{\pi}(M^2)$ 
to vanish at large $M^2$. Here we disagree with \cite{FMS} who claimed
that diffractive amplitudes are proportional to the integrated
gluon structure function $G(x,\bk^2)$. We notice, however, that in real 
photoproduction the cross section is dominated by the contribution 
from the helicity amplitude $\bPhi_{1}$ rather than $\Phi_0$ in
the pion case. Also, because of the pointlike $\gamma q\bar{q}$ 
QED vertex the photoproduction cross section is $\propto k^{-6}$,
see eq.~(29) of \cite{NZsplit}, compared to $k^{-8}$ for pions
as given by eq.~(\ref{eq:2.17}).
 
Second, in the same regime of hard dijets both diffractive 
amplitudes are proportional to the pion decay constant $F_{\pi}$
and, more important, to the pion distribution amplitude $\phi_{\pi}(z)$. 
By the nature of our derivation this property emerges if the 
radial wave function of the pion $\psi(z,\bk)$ is a steeper function 
of $\bk$ than $f^{(1)}(\bk)$, which holds naturally for the 
anticipated decrease of non-pointlike $\pi q\bar{q}$ vertex function 
and for the phenomenologically known gluon structure function of the 
proton. Consequently, the $z$-distribution of dijets allows the 
determination of the $z$-distribution of the pion distribution 
amplitude $\phi_{\pi}(z)$.

Third, we emphasize that to the leading twist the differential
cross section for dijet production on nucleons (eq.\ref{eq:2.17})
does not contain any free parameters, and thus is the perturbatively
calculable quantity.

%--------------  Section 3

\section{Nuclear diffraction amplitudes}

We consider coherent diffraction 
$$
\pi A \to q\bar{q} A'\,,
$$ 
where the 
recoil nucleus $A'$ remains in the ground state. We focus on the
forward diffraction cone $\bDelta^2 \lsim R_{A}^{-2}$, where $R_{A}$
is the nuclear radius. The longitudinal momentum transfer to the
nucleus equals $\Delta_{z} = x_{\Pom}m_{N}$ and coherent diffraction
is possible if $\Delta_{z}^2 \ll R_{A}^{-2}$, which condition is
satisfied in the E791 kinematics in which $x_{\Pom} \sim 10^{-2}$,
see also the discussion in section 4.1.

At $x_{\Pom} \sim 10^{-2}$ nuclear effects in DIS are dominated 
by nuclear shadowing of the  $q\bar{q}$ Fock state of the photon
\cite{NZ91,BGZ,Schaefer}. Hence  one must sum the $q\bar{q}$
multiple-scattering amplitudes of Fig.~3, we show a representative
set for the impulse approximation, $j=1$ (Figs. 3a,3b), and double scattering, 
$j=2$ (Figs. 3c-3e). 
The typical
multiplicity of rescatterings, $j$, is much smaller than the target 
mass number $A$. Because of the quark and antiquark helicity conservation 
one can calculate first the $q\bar{q}$-nucleus scattering amplitude
and convolute it with the pion wave function. Because the radius of
nuclei $R_{A}$ is much larger than the pion radius $R_{\pi}$ one can
safely neglect the $\bDelta$ dependence coming from the $q\bar{q}$-nucleon
scattering and take the $q\bar{q}$-nucleon amplitudes in
the forward limit $\bDelta=0$. The 
strong coupling enters the $q\bar{q}$ loop as $\alpha_{S}(\bk^2)$.
In the high-energy limit of $x_{\Pom}\ll 1/R_{A}m_{N}$ the 
calculation and summation of nuclear multiple scattering amplitudes
is readily done in the impact
parameter representation \cite{Glauber2,Gribov,NZ91,BGZ}. 
Namely, we notice that after passing to the 
$q\bar{q}$ color dipole representation the helicity amplitudes $\Phi_0(z,\bk)$ 
and $\bPhi_{1}(z,\bk)$ can
be cast in the form
\bea
\Phi_0 (z,\bfk) = \, \int d^2\bfr 
\, e^{\displaystyle -i\bfk \bfr} \, \sigma(x,\bfr)
\, m_{f}\Psi_\pi(\bfr , z) \nonumber \\
\bPhi_{1} (z,\bfk) = -i \, \int d^2\bfr 
\, e^{\displaystyle -i\bfk \bfr} \, \sigma(x,\bfr)
\, \bnabla \Psi_\pi(\bfr , z) \, .
\label{eq:3.1}
\eea
Here $\bfk$ is the transverse momentum of the jet, $\bfr$ is the $q\bar{q}$
separation in the impact parameter plane, 
\be
\Psi_{\pi}(z,\br)={1\over (2\pi)^2}\int d^{2}\bk \psi_{\pi}(z,\bk)
\exp(i\bk\br)
\label{eq:3.2}
\ee
is proportional to the $q\bar{q}$ color dipole distribution amplitude 
in the pion and
\bea
\sigma(x,\bfr) =  \alpha_{S}(\bk^2)\sigma_{0}
\int d^2 \bfkappa f^{(1)}(\bfkappa) \left[
1 - \exp( i \bfkappa \bfr)\right] \, 
\label{eq:3.3}
\eea
has the meaning of the dipole cross section
for interaction of the $q\bar{q}$ dipole $\br$ with the target nucleon
in which  the strong coupling $\alpha_S$ enters at the hard scale given by 
the jet transverse momentum $\bk$.

The Glauber-Gribov representation of amplitudes (\ref{eq:3.1}) for the nuclear
target is obtained by substitution of the $q\bar{q}$-nucleon scattering
amplitude by the $q\bar{q}$-nucleus scattering amplitude 
\cite{Gribov,NZ91,BGZ} and reads:
\bea
\Phi_0^{(A)}(z,\bfk,\bDelta)&=& 2m_{f}\, \int d^2\bfb \int d^2\br
e^{\displaystyle -i\bfb\bDelta-i\bfk \bfr} \Psi_\pi(z,\bfr)\nonumber
\left\{ 1- \exp\left[-{1\over2}\sigma(x,\bfr)T_A(\bfb)\right]\right\}
\nonumber \\
&=& 
2m_{f} \int d^2\bfb \int d^2\bfr e^{\displaystyle -i\bfb\bDelta-i\bfk \bfr} \Psi_\pi(z,\bfr) 
\, \sum_{n\geq 1} (-1)^{n+1}
{\dst \sigma^n (x,\bfr) \over \dst 2^{n} n!}  
T_A^{n} (\bfb)\, , \nonumber\\
\bPhi_{1}^{(A)}(z,\bfk,\bDelta)&=
&-2i\int d^2\bfb \int d^2\bfr e^{\displaystyle -i\bfb\bDelta-i\bfk \bfr} 
\left[\sum_{n\geq 1} (-1)^{n+1}
{\dst \sigma^n (x,\bfr) \over \dst 2^n n!} 
 T_A^n (\bfb)\right] \bnabla\Psi_\pi(z,\bfr)\,,
\nonumber\\
\label{eq:3.4}
\eea 
where $\bfb$ is the pion-nucleus impact parameter,
 $T_A(\bfb)=\int dz' n_{A}(\bfb,z')$ is the familiar nuclear optical thickness
\cite{Glauber2},
$n_A(\bfb,z')$ is the nuclear matter density. The frozen dipole approximation 
(\ref{eq:3.4}) is applicable because $\Delta_{z}^2 \ll R_{A}^{-2}$ (\cite{Gribov},
for the modern formalism see \cite{BGZ}).
For the sake of simplicity above we took the exponentiated form
for the nuclear profile function instead of its more exact form
\bea
\Gamma_A (\bfb) &=& 
1 - \left[ 1 - {\sigma(x,\bfr) T_A(\bfb) \over 2 A} \right]^A \nonumber \\
&\simeq&  \left[ 1- \exp\left[-{1\over2} \sigma(x,\bfr)T_A(\bfb)\right]\right] \, ,
\label{eq:3.5}
\eea
reformulation of all results for the polynomial form poses no problems.

Representative diagrams for the impulse approximation, $j=1$, are shown
in Figs.~3a, 3b. They give the familiar result
\be
\Phi_0^{(A)}(z,\bfk,\bDelta)=\Phi_0(z,\bfk)\int  d^2\bfb e^{\displaystyle 
-i\bfb\bDelta}T_A(\bfb)=
A\Phi_0(z,\bfk)G_{em}(\bDelta)\, ,
\label{eq:3.6}
\ee
where $G_{em}(\bDelta)$ is the charge form factor of the nucleus.

The principal effect of rescattering is readily seen from the double
scattering, $j=2$. Making use of the integral representation (\ref{eq:3.3}) we find
\bea
\int d^2\bfr e^{\displaystyle -i\bfk \bfr} \Psi_\pi(\bfr,z) \sigma^2 (x,\bfr) &=&
\alpha_S^2(\bk^2)\sigma_{0}^2
\int  d^2\bfkappa_1 d^2\bfkappa_2 f^{(1)}(\bfkappa_1) f^{(1)}(\bfkappa_2) \nonumber \\ 
&\times&  \int d^2\bfr \, e^{\displaystyle -i\bfk\bfr} 
\left[1 -  e^{\displaystyle i\bfkappa_1\bfr} - e^{\displaystyle i\bfkappa_2\bfr}
 + e^{\displaystyle i(\bfkappa_1+\bfkappa_2)\bfr} \right] \Psi_\pi(\bfr,z)  \nonumber \\
 &=&  \alpha_S^2(\bk^2)\sigma_{0}^2
\int  d^2\bfkappa_1 d^2\bfkappa_2 f^{(1)}(\bfkappa_1) f^{(1)}(\bfkappa_2) 
\nonumber \\ 
&\times&  \left[ \Psi_\pi(\bfk,z) - 2\,\Psi_\pi(\bfk - \bfkappa_1,z) + 
\Psi_\pi(\bfk - \bfkappa_1 - \bfkappa_2,z) \right] 
\nonumber \\ 
&=& \alpha_S^2(\bk^2)\sigma_{0}^2
\left[\Psi_\pi(\bfk,z)  - 2 \int
d^2\bkappa \Psi_\pi(\bfkappa,z)f^{(1)}(\bk -\bfkappa) \right.\nonumber\\
&+&\left. 
\int d^2\bkappa \Psi_\pi(\bkappa,z) f^{(2)}(\bk -\bfkappa)\right]
\, .
\label{eq:3.7}
\eea
where
\be
f^{(2)}(\bk)=\int  d^2\bfkappa_{1} d^2\bfkappa_{2}f^{(1)}(\bfkappa_{1}) 
f^{(1)}(\bfkappa_2)\delta(\bk -\bfkappa_{1}-\bfkappa_{2}) 
\label{eq:3.8}
\ee
is normalized to unity: $\int d^2\bk f^{(2)}(\bk)=1.$
The sum of the impulse approximation, $n=1$, and double-scattering terms
equals (for the sake of illustration we take $\bDelta=0$):
\bea
\Phi_0^{(A)}(z,\bk,\bDelta=0)=\alpha_{S}(\bk^2)\sigma_{0}m_{f}
\int d^2\bfb T_{A}(\bfb) \left\{
\psi_{\pi}(z,\bk)\left[1- {1\over 2}\alpha_{S}(\bk^2)\sigma_{0}T_{A}(\bfb)\right]
\right. \nonumber\\
-\int d^2\bkappa 
\psi_{\pi}(z,\bkappa)f^{(1)}(\bk -\bkappa)
\left[1- {1\over 2}\alpha_{S}(\bk^2)\sigma_{0}T_{A}(\bfb)\right] \nonumber\\
-\left. {1\over 2}\int d^2\bkappa 
\psi_{\pi}(z,\bkappa)f^{(2)}(\bk -\bkappa)
\alpha_{S}(\bk^2)\sigma_{0}T_{A}(\bfb)\right\}\,.
\label{eq:3.9}
\eea
The three terms in the last line of (\ref{eq:3.7}) and in (\ref{eq:3.9})
correspond to the three classes of double scattering diagrams shown in 
Figs. 3c-3e. The first term in
the r.h.s. of (\ref{eq:3.9}) shows that the no-Pomeron splitting term
in (\ref{eq:2.12}) coming from Fig.~3a in the nuclear case 
receives the conventional shadowing correction 
from the double-scattering diagram of Fig. 3c (and the not shown here 
partner diagram
in which the two gluons from second nucleon couple
to the antiquark). The Pomeron splitting
term in (\ref{eq:2.12}) coming from Fig.~3b in the nuclear case
is similarly shadowed by double-scattering
diagrams of Fig. 3d. The effective
shadowing cross section equals 
\be
\sigma_{eff}(\bk^2)=\alpha_{S}(\bk^2)\sigma_{0}
\label{eq:3.10}
\ee
The new feature of double scattering is the  third term in ({\ref{eq:3.9}) given by
the double-Pomeron splitting diagram of Fig.~3e. The convolution (\ref{eq:3.8})
implies the broadening of $f^{(2)}(\bk)$ compared to $f^{(1)}(\bk)$. Furthermore,
this broadened distribution $f^{(2)}(\bk)$ has the same sign as, i.e. it is
an antishadowing correction to, the single-Pomeron splitting term (\ref{eq:3.9})
and would eventually take over single-Pomeron splitting at large $\bk$. 

Higher order rescatterings give rise to distributions with $j$-fold Pomeron
splitting 
\be
f^{(j)}(\bk)=\int  d^2\bfkappa_{1}... d^2\bfkappa_{j}f^{(1)}(\bfkappa_{1}) 
f^{(1)}(\bfkappa_2)...f^{(1)}(\bfkappa_j)\delta(\bk -\sum_{i=1}^{j} 
\bfkappa_{i}) 
\label{eq:3.11}
\ee
which obviously broaden with increasing $j$.
Rearranging nuclear diffractive amplitudes as an expansion over $f^{(j)}(\bk)$
we obtain 
\bea
\Phi_{0}^{(A)}(z,\bfk) &=& 2  m_{f}\, \sum_{j\geq 1} \int d^2\bfb e^{\displaystyle -i\bfb \bDelta} 
\int d^2\bkappa \left[ \Psi_\pi(z,\bfk) - \Psi_{\pi}(z,\bfk-\bkappa) \right]
\,f^{(j)}(\bkappa) \,\nonumber\\
&\times& {1\over j!}
 \left[{\sigma_{eff}(\bk^2)T_A(\bfb)\over 2}\right]^j 
\exp \left[ -{\sigma_{eff}(\bk^2)\over 2} T_A(\bfb) \right] \nonumber\\
&=& 2  m_{f}\, \int d^2\bfb e^{\displaystyle -i\bfb \bDelta} 
 \left\{ \Psi_\pi(z,\bfk) 
\left[ 1 - \exp \left( -{\sigma_{eff}(\bk^2)\over 2}  T_A(\bfb) \right)\right] 
\right. \nonumber \\ 
&-& \left. \sum_{j\geq 1}  
   \int d^2\bkappa \Psi_{\pi}(z,\bkappa) 
f^{(j)}(\bfk-\bkappa){1\over j!}
 \left[{\sigma_{eff}(\bk^2)T_A(\bfb)\over 2}\right]^j 
\exp \left[ -{\sigma_{eff}(\bk^2)\over 2}  T_A(\bfb) \right] \right\} \, ,
\nonumber\\
\bPhi_{1}^{(A)}(z,\bfk) &=& 2\, \sum_{j\geq 1} \int d^2\bfb e^{\displaystyle -i\bfb \bDelta} 
\int d^2\bkappa \left[\bfk \Psi_\pi(z,\bfk) - 
(\bfk-\bkappa)\Psi_{\pi}(z,\bfk-\bkappa) \right]
\,f^{(j)}(\bkappa) \,\nonumber\\
&\times& {1\over j!}
 \left[{\sigma_{eff}(\bk^2)T_A(\bfb)\over 2}\right]^j 
\exp \left[ -{ \sigma_{eff}(\bk^2)\over 2} T_A(\bfb) \right] \nonumber\\
&=& 2 \, \int d^2\bfb e^{\displaystyle -i\bfb \bDelta} 
 \left\{ \bfk\Psi_\pi(z,\bfk) 
\left[ 1 - \exp \left( -{\sigma_{eff}(\bk^2)\over 2} T_A(\bfb) \right)\right] 
\right. \nonumber \\ 
\!&-&\! \left. \sum_{j\geq 1}  
   \int d^2\bkappa \bkappa\Psi_{\pi}(z,\bkappa) 
f^{(j)}(\bfk-\bkappa){1\over j!}
 \left[{\sigma_{eff}(\bk^2)T_A(\bfb)\over 2}\right]^j 
\exp \left[ -{\sigma_{eff}(\bk^2)\over 2}  T_A(\bfb) \right] \right\} \, .
\nonumber\\
\label{eq:3.12}
\eea 
The nuclear attenuation factors show that shadowing is indeed controlled by 
$\sigma_{eff}(\bk^2)$. Despite the decrease  
$\sigma_{eff}(\bk^2) \propto \alpha_{S}(\bk^2)$
numerically this cross section 
is quite large, grows slowly at very small $x_{\Pom}$, and
is a soft gluon exchange dominated quantity.

At large jet momentum the diffractive amplitude is dominated by the 
second term in (\ref{eq:3.12}) in which all broadened $j$-Pomeron 
splitting contributions enter remarkably with the same antishadowing sign.
 
Whether this antishadowing takes over shadowing depends on the large $\bk^2$ 
behaviour of $f^{(2)}(\bk)$. For the power asymptotics (\ref{eq:2.12}) the 
leading contribution to convolution $f^{(2)}(\bk)$ (\ref{eq:3.8}) at large 
$\bk^2$ comes from the configurations when there is one hard splitting of 
the Pomeron with $\kappa_{i}\sim \bk$, whereas all other $\kappa_{i}$ are 
small. Correspondingly, the values of gluon lightcone momenta $x_{i}$ in 
the hard splitting of the Pomeron are the same as in the free nucleon case, 
whereas in the predominantly soft rescatterings $x_{i} \sim x_{S} \sim 
1 {\rm GeV}^2/W^2$ and in multiple scattering expansion 
(\ref{eq:3.12}) all $\sigma_{eff}(\bk^2)$ but one for the hard splitting 
of the Pomeron must  arguably be evaluated at $x=x_{S}$. However, we 
notice that according to \cite{INDGSF} the $x$-dependence of 
${\cal F}(x,\bk^2)$ is weak for soft $\bk^2$ and in the practical evaluation of 
$f^{(j)}(\bk)$ one can put $x_{S}=x_{\Pom}/2$. 
Then to the leading twist one readily finds the large-$\bk^2$ asymptotics
\be
f^{(j)}(\bk) = jf^{(1)}(\bk)
\label{eq:3.13}
\ee
which clearly shows the anticipated enhancement of the large-$\bk^2$
tail by multiple rescatterings. The salient feature of
large-$\bk^2$ broadening
(\ref{eq:3.13}) is that the exponent of the power asymptotics
is sustained. Because of the normalization condition
$\int d^2\bk f^{(j)}(\bk)=1$ the broadening at large $\bk$ entails
the small-$\bk$ depletion $f^{(j\geq 2)}(\bk) <  f^{(1)}(\bk)$ at
small-$\bk$. Evidently, for larger $j$ this small-$\bk$ depletion 
will be stronger and would extend to larger $k$. 

The evaluation of the higher twist correction making
use of the expansion (\ref{eq:2.14}) proceeds
as follows. For the sake of definiteness focus on the contribution
from the configuration in which $\bkappa_{1} \approx \bk$ and the total 
transverse momentum for the subset 
$i=2,...,j$ is small, 
$(\sum_{i\geq 2}\bkappa_{i})^2 \lsim \bk^2$.
 Then, to the Leading Log$\bk^2$ accuracy the higher twist contribution
will be dominated by the $(j-1)$ configurations of the subset 
$i=2,...,j$ in which one of the $\bkappa_{i}^2$ 
is running up to $\sim \bk^2$ whereas all other momenta are small:
\bea
f^{(j)}(\bk) \simeq jf^{(1)}(\bk) 
\int  d^2\bfkappa_{2}... d^2\bfkappa_{j}
\left[1+{\delta^2 \over \bk^2} (\sum_{i\geq 2}\bkappa_{i})^2\right]
f^{(1)}(\bfkappa_2)...f^{(1)}(\bfkappa_j) \nonumber\\
=jf^{(1)}(\bk)\left[1+{\delta^2 \over \bk^2}(j-1)\int^{\bk^2} d^2\bkappa
\bkappa^2 f^{(1)}(\kappa)\right]
= jf^{(1)}(\bk)\left[1+{4\pi^2\delta^2 \over 3\sigma_{0}\bk^2}(j-1)
G({1\over 2}x_{\Pom},\bk^2)
\right]
\label{eq:3.14}
\eea
Remarkably, the coefficient of the higher twist correction is proportional to the 
gluon structure function of the proton $G({1\over 2}x_{\Pom},\bk^2)$ 
at a hard scale $\bk^2$.
For the $j$-Pomeron splitting diagrams the higher twist correction rises 
$\propto (j-1)$, i.e. the antishadowing/broadening of large-$\bk^2$ tail 
takes place also to the higher twist. 

Making use of the result (\ref{eq:3.14}) in the expansion ({\ref{eq:3.12})
we obtain
\bea
\Phi_{0}^{(A)}(z,\bfk,\bDelta) = 2 \, \int d^2\bfb e^{\displaystyle -i\bfb \bDelta} 
 \left\{ m_{f}\Psi_\pi(\bk,z) 
\left[ 1 - \exp \left( -{1\over 2} \sigma_{eff}(\bk^2) T_A(\bfb) \right)\right] 
\right. \nonumber \\ 
- \left.     
{1\over 2}f^{(1)}(\bfk)F_{\pi}\phi_{\pi}(z)\sigma_{0}\alpha_{S}(\bk^2)
\left[T_A(\bfb)+ {2\pi^2\delta^2 \alpha_{S}(\bk^2)\over 3 \bk^2}G(\bk^2)
T_{A}^{2}(\bfb) \right]\right\}  \, .
\label{eq:3.15}
\eea 
Here the impulse approximation term $\propto \Psi_{\pi}(\bk,z)$ is shadowed 
with the soft cross section (\ref{eq:3.10}). 
The most remarkable feature of the Pomeron splitting contributions 
(\ref{eq:3.15}) is an exact cancelation of soft shadowing and 
antishadowing/broadening effects. Furthermore, this 
cancelation makes redundant the
exact value of $x_{S}$ at which $\sigma_{0}$ must be taken in the 
nuclear multiple scattering expansion. The broadening law (\ref{eq:3.13}) 
is crucial for this cancelation of soft shadowing and 
antishadowing/broadening effects. Similar exact cancelation
of shadowing and antishadowing/broadening effects, and 
independence on the exact value of soft shadowing cross section
$\sigma_{0}$, take place in the next-to-leading twist correction too. 
Consequently, both leading and next-to-leading twist amplitudes
are parameter free calculable in hard perturbative QCD.
 
Following the derivation (\ref{eq:2.16}) for the free nucleon target, a 
similar analysis can be repeated for the nuclear target with the result
\bea
\bPhi_{1}^{(A)}(z,\bfk,\bDelta) &=& 2 \bk
\, \int d^2\bfb e^{\displaystyle -i\bfb \bDelta} 
 \left\{ \Psi_\pi(\bfk,z) 
\left[ 1 - \exp \left( -{1\over 2} \sigma_{eff}(\bk^2) T_A(\bfb) \right)\right] 
\right. \nonumber \\ 
&-& \left.{1\over 2} \sigma_{0}\alpha_{S}(\bk^2) 
   {\delta \langle \kappa_{\pi}^2(z)\rangle  
\over m_{f} \bk^2}f^{(1)}(\bk)F_\pi \phi_{\pi}(z) 
T_A(\bfb)\right\}  \, ,
\label{eq:3.16}
\eea 
where we neglect the corrections $\propto k^{-2}$ to the already higher
twist convolution term.

We notice that within the diffraction cone of $\bDelta^2 \lsim R_{A}^2$ 
the viable approximation is
\be
\int d^2\bfb e^{\displaystyle -i\bfb \bDelta}T_A^{2}(\bfb)=
{3C_{A} A^2 \over 4\pi \langle R_{ch}^2 \rangle } G_{em}({1\over 2}\bDelta^2)\,,
\label{eq:3.17}
\ee
where the coefficient $C_{A}\approx 1$ depends slightly on the shape
of the nuclear matter distribution: $C_{A}=1$ for the Gaussian density 
appropriate for light nuclei decreases slowly to $C_{A}={9\over 10}$
for the uniform density sphere. Then for large jet momentum $\bk$ when the 
impulse approximation contribution dies out, our nuclear diffractive amplitudes 
(\ref{eq:3.15}), (\ref{eq:3.16}) take a particularly simple form
\bea
\Phi_0^{(A)}(z,\bfk,\bDelta) = - A F_{\pi}\phi_{\pi}(z) 
{4 \pi \alpha_{S}(\bk^2)\over 3}{{\cal{F}}({1\over 2}x_{\Pom},\bk^2) \over \bk^4}
\left(1+\delta^2 {\langle \kappa^2_\pi(z)\rangle \over \bk^2}
\right)\nonumber\\
\times
\left\{ G_{em}(\bDelta^2)
+ {\pi\delta^2 C_{A}A\alpha_{S}(\bk^2)\over 2\langle R_{ch}^2\rangle \bk^2} 
G({1\over2}x_\Pom,\bk^2)G_{em}({1\over 2}\bDelta^2)
\right\}  \, ,
\label{eq:3.18}
\eea 
\bea
\bPhi_{1}^{(A)}(z,\bfk,\bDelta) &=& -\bk\,A 
F_\pi \phi_{\pi}(z) {4 \pi \alpha_{S}(\bk^2)\over 3}
{{\cal{F}}({1\over 2}x_{\Pom},\bk^2) \over \bk^4}G_{em}(\bDelta^2)
{\delta\langle \kappa^2_\pi(z)\rangle  
\over  m_{f} \bk^2} \, .
\label{eq:3.19}
\eea 

Finally, a correction for the finite longitudinal momentum transfer to
the nucleus can be evaluated as follows. First, 
within the diffraction cone the impulse approximation
amplitude (\ref{eq:3.6}) acquires the longitudinal form factor 
$G_{em}(x_{\Pom}^2m_{N}^2)$. Second, we have shown that in multiple 
rescattering contributions there is only one hard splitting of the
Pomeron with the longitudinal momentum transfer $\Delta_{z}
\approx x_{\Pom}m_{N}$ whereas in soft rescatterings the longitudinal
momentum transfer can be neglected. As a result, the same longitudinal 
form factor $G_{em}(x_{\Pom}^2m_{N}^2)$ holds for all  multiple 
rescatterings.
 
Upon the $\bDelta^2$ integration we find that the large-$\bk^2$ asymptotics of 
nuclear diffraction cross section will read
\bea
{d\sigma_D \over dz d\bk^2 }= 
{2\pi^5 \over 27} F^2_{\pi} \phi^2_{pi}(z) 
G_{em}^{2}( x_{\Pom}^2m_{N}^2) \alpha^2_S(\bk^2)
\left[{{\cal{F}}({1\over 2}x_{\Pom},\bk^2) \over \bk^4}\right]^2
\cdot {3A^{2} \over \langle R_{ch}^2\rangle }\nonumber\\
\cdot\left\{ 1 +  {\delta^2 \over \bk^2}\left[
2\langle \kappa^2_\pi(z)\rangle\left( 1+ {\langle \kappa^2_\pi(z)
\rangle \over 2m_{f}^2}\right)
+ {2\pi C_{A}A\alpha_{S}(\bk^2)\over 3\langle R_{ch}^2\rangle }
 G({1\over 2}x_{\Pom},\bk^2)\right]\right\}\, ,
\label{eq:3.20}
\eea
where we included the effect of the longitudinal form factor.
Clearly, at a sufficiently large $k$ the higher twist correction will be
dominated by the last term which is model-independent, enhanced by the 
gluon structure function $G({1\over 2}x_{\Pom},\bk^2)$ and rises for heavy nuclei 
$\propto A^{1\over 3}$.

%-----------  section 4
 
\section{The numerical results and comparison with the E791 data}

% --------   section 4.1

\subsection{Nuclear broadening of multiple-Pomeron splitting 
contributions}

In the E791 kinematics $1.25$ GeV$\lsim k \lsim 2.5$ GeV and $W^2 = 940$ GeV$^2$, 
so that $x_{\Pom} \sim $(1--2)$\cdot 10^{-2}$.
We start with a check of the accuracy of the expansion (\ref{eq:3.14})
according to which 
\be
\Delta_{j}(k) = {1 \over j-1} \left( {f^{(j)}(\bk) \over j f^{(1)}(\bk)} -1\right)
\approx {2\pi^2\delta^2 \over 3\sigma_{0} \bk^2} G({1\over 2}x_{\Pom},\bk^2)
\label{eq:4.1.1}
\ee
must exhibit $j$-independence at sufficiently large $\bk^2$.  
The results for $x_{\Pom}=2\cdot 10^{-2}$ shown in Fig.~4a demonstrate
this is indeed the case for very large $k \gsim$ (2--3) GeV. Slight departure from
the universality can be understood in terms of the slight $k$ dependence
and the broadening driven $j$ dependence of the effective 
exponent $\delta$.  The situation at lower $k$ is 
a nontrivial one, because at a sufficiently small $k$ the broadening 
of $f^{(j)}(\bk)$ must be superseded by the small-$\bk$ depletion, in which
region 
\be
\Delta_{j}(k) \sim -{1\over j}\, .
\label{eq:4.1.2}
\ee
The effect of depletion extends to larger $k$ with increasing $j$. Indeed, 
for $k=1$ GeV and $j=10$ our $\Delta_{j}(k)$ is getting close to the 
no-broadening estimate (\ref{eq:4.1.2}). Anyway, the finding of 
$\Delta_{j}(k) < 0$ indicates breaking of the broadening law (\ref{eq:3.13})
for $j \gsim 5 $ at $k\sim 1$ GeV, the point of crossover 
$\Delta_{j}(k)=0$  is moving to larger $k$ with increasing $j$. 
The large-$k$ 
asymptopia is even more elusive at smaller $x$, see Fig.~4b for 
$x_{\Pom}=2\cdot 10^{-3}$. Here the broadening law (\ref{eq:3.13})
is only applicable at $k\gsim $ (4--5) GeV.

According to these results, the E791 range of $k$ falls in the transient 
region in which the higher twist expansion (\ref{eq:3.14}) is not applicable yet
and the multiple-scattering broadening is weaker than given by (\ref{eq:3.13}). 
This suggests that in this transient region of $k$ the large-$k$ 
broadening of $f^{(j)}(\bk)$ is not sufficient for exact cancelation of 
shadowing effects. All numerical estimates of the A-dependence must use
exact numerical results for $f^{(j)}(\bk)$.

%---------  section 4.2

\subsection{The pion wave function and z-distribution amplitude}

In numerical calculations of diffraction amplitudes we use a slight 
modification of the Jaus \cite{Jaus} parameterization of the pion wave 
function, which in our convention (\ref{eq:2.10}) is  
\be
\psi_{\pi}(z,\bk) \propto {1 \over z(1-z)M^2}
\exp\left(-{1\over 8}R_{\pi}^{2}
(M^{2}-4m_{f}^2)\right)
\label{eq:4.2.1}
\ee
and with $R_{\pi}= 2.2$ GeV$^{-1}$ and $m_{f}=m_{u,d}=0.215$ GeV provides 
a consistent description of the $\pi \to \mu \nu$ decay constant $F_{\pi}$,
charge radius of the pion, $\pi^{0}\to 2\gamma$ decay rate and slope of the 
$\pi^{0}\to \gamma\gamma^{*}$ form factor 
\cite{Schwiete} (for the related analysis see \cite{Jaus}). 
The numerical results for $\phi_{\pi}(z)$ are shown in Fig.~5 and 
differ only weakly from the often discussed  
asymptotic distribution amplitude $\phi_{asym}(z)=6z(1-z)$ (for 
the review see \cite{CZ,Kroll,Brodsky}), in the broad range $0.2 \lsim z \lsim 0.8$
the difference does not exceed 10\%. A convenient analytic approximation 
to $\phi_{\pi}(z)$ given by this soft wave function 
is 
\be
\phi_{\pi}(z)=  0.6572 \log \left(1+{8z(1-z)\over R_{\pi}^{2}m_{f}^2}\right)
\exp\left(-{R_{\pi}^{2}m_{f}^2 \over 8z(1-z)}\right)
\label{eq:4.2.2}
\ee
and is good to better than 1 per cent apart from $z\lsim 0.03$ and 
$1-z \lsim 0.03$.

The large-$\bk$ asymptotics of the helicity amplitude for excitation 
of dijets with $\lambda+\bar{\lambda}=\pm 1$ is proportional to 
$\langle \kappa^2_\pi(z)\rangle$, defined in (\ref{eq:2.15}). The soft
wave function (\ref{eq:4.2.1}) gives
\be
\langle \kappa^2_\pi(z)\rangle \approx { 8z(1-z) \over  R^2 \log\left(
1+ {\dst 8z(1-z) \over \dst R^2m_{f}^2}\right)}\, ,
\label{eq:4.2.3}
\ee  
i.e., $\langle \kappa^2_\pi(z=0.5)\rangle \approx 0.17$ GeV$^2$, which is
a natural scale for the soft pion wave function. It decreases gradually
away from $z=0.5$, for instance $\langle \kappa^2_\pi(z=0.2)\rangle =
\langle \kappa^2_\pi(z=0.8)\rangle \approx 0.12$ GeV$^2$.

Here we notice that in view of (\ref{eq:2.1}) and (\ref{eq:2.4}) 
there is the kinematical $z$-$x_{\Pom}$ correlation $x_{\Pom} \propto
[4z(1-z)]^{-1}$. The $x_{\Pom}$ dependence of DGSF can be parameterized as
${\cal F}({1\over 2}x_{\Pom},\bk^2) 
\propto x_{\Pom}^{{\displaystyle \tau(\bk^2)}}$.
With allowance for the $z$-$x_{\Pom}$ correlation the observed 
$z$-dependence of diffractive amplitudes changes from $\phi_{\pi}(z)$ 
to 
\be
 \phi_{\pi}^{obs}(\tau,z) \propto 
\phi_{\pi}(z)[z(1-z)]^{{\displaystyle \tau(\bk^2)}}\, .
\label{eq:4.2.4}
\ee
The phenomenological determination of the exponent $\tau(\bk^2)$ 
in \cite{INDGSF} gave $\tau \approx 0.16$ at $k=1.35$ GeV and 
$\tau \approx 0.22$ at $k=2$ GeV for the GRV-D and MRS-D parameterizations
for ${\cal F}({1\over 2}x_{\Pom},\bk^2)$ and $\tau \approx 0.25$ at 
$k=1.35$ GeV and $\tau \approx 0.30$ at $k=2$ GeV for the CTEQ-D
parameterization. In Fig.~5 we show the observed z-distribution
amplitudes for $\tau=0.15$ and $\tau=0.30$, evidently the observed 
$z$ distribution gets even closer to $\phi_{asym}(z)$.

% ------------  section 4.3 

\subsection{Importance of multiple Pomeron splitting processes 
in nuclear diffraction}

The result for the Pomeron splitting term in (\ref{eq:3.15}) has a deceptively 
simple form of the sum of the single and double scattering terms but such an 
interpretation would be utterly wrong. 
An importance of multiple nuclear
rescatterings can be judged from the mean value of $j$ in expansion 
(\ref{eq:3.12}). Casting it in the form $\Phi_0^{(A)}(z,\bk)= 
\sum_{j} w_{j}f^{(j)}(\bk)$, for $k=$2 GeV we find
\be
 \langle j \rangle = {\sum_{j} w_{j}jf^{(j)}(\bk)\over \sum_{j} w_{j}f^{(j)}(\bk)}
=\left\{\begin{array} {rl}2.03 & \mbox{\rm for $^{12}C$}\\
4.10 & \mbox{ for  $^{196}Pt$} ,\nonumber \end{array}\right.
\label{eq:4.3.1}
\ee
which shows clearly an inadequacy of truncation of nuclear rescatterings 
to the single and double scattering. Indeed, at $x\sim 10^{-2}$ and 
$\bk^{2}\sim 4$ GeV$^2$ the effective shadowing cross section is
quite large, $\sigma_{eff}(\bk^2) \sim 40$ mb. Closer inspection shows that for
exhausting  95 \% of the strength of $\Phi_0^{A}$ one needs the 
contributions up to $j=4$ for the carbon target (A=12) and up to 
$j=$(8--9) for the platinum target (A=196). For the E791 
energy, i.e. $x_{\Pom} \sim 10^{-2}$, this implies in conjunction
with the results shown in Fig.~4 that from the viewpoint of convergence
of nuclear expansions $k \gsim $ 2.5--3 GeV are
needed for the applicability of the leading plus next-to-leading 
expansions (\ref{eq:3.18}), (\ref{eq:3.19}), (\ref{eq:3.20}).

% ----------   section 4.4

\subsection{Evaluation of next-to-leading twist corrections}

The next-to-leading twist correction can conveniently by
parameterized as $H/k^2$.
First we evaluate the contribution to $H$ coming from 
the pion wave function. It is controlled by the moment
$\langle \kappa^2_\pi(z)\rangle$. For the $0.2 \lsim z \lsim 0.8$
which are relevant to the E791 data, we take in further estimates 
$\langle \kappa^2_\pi(z)\rangle = 0.15$ GeV$^2$. We notice that 
numerically $\langle \kappa^{2}(z=0.5)\rangle \approx 3.5 m_{f}^2$, 
which entails that on the free nucleon target next-to-leading 
twist effects come predominantly from the helicity component 
$\propto \bPhi_1^2$ in the cross section, see eq.~(\ref{eq:2.17}):  
\be 
H_{\pi}^{(1)} ={\delta^2\langle \kappa^2_\pi(z)\rangle^{2} \over m_{f}^2} \approx 
0.55 \delta^2 \, {\rm GeV}^{-2} \sim 2.2 \, {\rm GeV}^{-2}\, .
\label{eq:4.4.1}
\ee
Here we took $\delta^2 \sim$ 4 appropriate for $k\gsim $ (2.5--3) 
GeV where the asymptotic expansion (\ref{eq:3.20}) is 
applicable. 
 
A similar estimate for the contribution to $H$ from from the 
helicity component $\propto \Phi_0^2$ in the cross section is
\be
H_{\pi}^{(0)} \sim 1.3 \, {\rm GeV}^{-2}\,.
\label{eq:4.4.2}
\ee

For $x_{\Pom} \sim 10^{-2}$ and $\bk^2\sim $ (3--4) GeV$^{2}$ relevant
to the E791 kinematics, the analysis \cite{INDGSF} gives 
$\alpha_{S}G({1\over 2}x_{\Pom},\bk^2)\approx$ 1--1.2. Then the nuclear 
rescattering/broadening contribution to next-to-leading twist is
(we use the nuclear density parameters from the compilation \cite{Adata})

\be
H_{A} = {4\pi \delta^{2}C_{A}A^{2}\over 3\langle R_{ch}^2\rangle }
\alpha_{S}(\bk^2)G({1\over 2}x_{\Pom},k^2)
\approx (0.16-0.2)\delta^2 A^{1\over 3} \sim 
\left\{\begin{array} {rl} 1.5 \, \mbox{GeV$^2$} & \mbox{\rm for $^{12}C$}\\
3.5 \, \mbox{GeV$^2$} & \mbox{ for  $^{196}Pt$} ,\nonumber \end{array}\right.
\, ,
\label{eq:4.4.3}
\ee
Lumping together all three contributions, we find $H^{Pt} \sim 7 $ 
GeV$^2$  and $H^{C} \sim 5$ GeV$^2$. Consequently, no simple expansion
in the leading and next-to-leading contributions is possible 
in the E791 region of jet momentum and an accurate numerical 
evaluation of broadening of $f^{(j)}(\bk)$ is called upon.

%-------   section 4.5

\subsection{Comparison with the E791 data: $\bk^2$ distributions}

In Fig.6 we show our numerical results for the $\bDelta^2$-integrated 
nuclear diffraction cross section for the platinum target. 
We are interested in the large $\bk$ where the contributions 
$\propto \psi_{\pi}(z,\bk)$ died out, 
which in both helicity amplitudes is preceded by the zero.
In order to have a crude idea on where this happens, we stretched 
our calculations for platinum target down to $k=0.5$ GeV. The 
amplitude $\Phi_0$ has a zero at $k\sim 0.65$ GeV. Because the term
$\propto \psi_{\pi}(z,\bk)$ is sensitive to nuclear shadowing 
and to the soft cross section $\sigma_{0}$ thereof, the position of
the zero is model-dependent. In $\bPhi_{1}$ the second component is
of higher twist and $\bPhi_{1}$ has a zero at larger $k \sim 1 $ GeV.
The impact of these zeros of diffractive amplitudes is manifest
up to $k\gsim 1.5$ GeV, which is still another reason why a 
comparison of our predictions with the experimental data
is justified only at $k\gsim 1.5$ GeV.

The E791 data give only the $k$ dependence of the acceptance-corrected 
cross section without absolute normalization. The normalization 
of our theoretical curve is the eyeball fit to the data, the agreement 
with the experimentally observed $k$-dependence is good at $k \gsim 1.5$
GeV. In the theoretical calculations we 
include consistently the $\bk^2$-$x_{\Pom}$ correlation discussed
in section 2 following eq. (\ref{eq:2.17}). 

At E791 energy our leading twist theoretical cross section 
would have followed 
the law $k^{-n}$ with the slope 
\be
n\approx 4+2\delta + 2\tau(\bk^2) \approx  8.7-8.8\, . 
\label{eq:4.5.1}
\ee
Our numerical results do not exhibit simple dependence 
$\propto k^{-n}$, if we define the local slope  $n$  as
$$
n = -{ \partial \log \sigma_{D} \over \partial \log \bk^2}\, ,
$$
then for $k\gsim 2.5$ GeV we find $n\sim 12$, whereas around 
$k=2$ GeV the slope $n\approx 10$. We attribute these large values
of local $n$  to very large higher twist corrections.

In Fig.~6 and Fig.~7 we show separately the contributions to the diffractive cross
section of the two helicity states: the leading plus next-to-leading 
twist component with $\lambda+\bar{\lambda}=0$ and the higher twist
component $\lambda+\bar{\lambda}=\pm 1$. They are of comparable 
magnitude, the higher twist component $\propto\bPhi_{1}^2$ starts
dying out and the component $\propto \Phi_0^2$ starts taking over
only at $k\gsim$ 2.5--3 GeV, in perfect agreement with 
evaluation of higher twists in section 4.4.

Fig. ~7 shows our predictions for the future experimental tests 
of the absolute normalization of diffractive cross section.
In all the numerical calculations we used the parametrization
labeled 'D-GRV' from \cite{INDGSF}. Here for the 
sake of convenience we plot $k^{8}d\sigma_{D}/dz dk^2|_{z=0.5}$.
As an illustration of the energy dependence of  dijet production
here we also show the predictions for $E_{\pi}=5$ TeV. In this
case $x_{\Pom} \sim 10^{-3}$ and $\delta \sim 1.7$, so that in the
leading twist we expect $n\sim 8$. Furthermore, because of the lower
value of $\delta$ the higher twist effects would be about half of those 
for $E_{\pi}=500$ GeV in the E791 experiment, cf. Fig.~4a and 4b. 
Indeed, for $E_{\pi}=5$ TeV  
$k^{8}d\sigma_{D}/dz dk^2|_{z=0.5}$ exhibits much weaker dependence 
on $\bk^2$.

\subsection{Comparison with the E791 data: nuclear mass number dependence}
 
The $A$-dependence of nuclear cross sections is often parameterized 
as $\sigma \propto A^{\alpha}$. The E791 experiment uses the carbon 
and platinum targets and defines the exponent $\alpha$ as
$$
\alpha = { \log \displaystyle{\sigma_{1} \over \sigma_{2}} \over 
\log\displaystyle {A_{1} \over A_{2}}}\, .
$$ 
For the reference, in the impulse approximation 
$$
\sigma \propto {A^2 \over \langle R_{ch}^2\rangle}
$$
and simple evaluation for Pt and C nuclei gives $\alpha_{IA}=1.44$ 
(the average slope for the $A$ dependence in the same interval gives
$\alpha_{IA}=1.39$, for instance, see \cite{KNNZCT}). A comparison 
of our numerical results for Pt and C nuclei yields the $k$-dependence 
of the exponent $\alpha$ shown in Fig.~8.

As we discussed in section 4.1, at $k\lsim 1.5$ GeV antishadowing/broadening
effects are too weak to cancel nuclear shadowing, which explains 
the small value of $\alpha(\bk^2) < \alpha_{IA}$. On the other hand, 
at larger $k$ higher-twist corrections become substantial and 
give $\alpha(\bk^2) > \alpha_{IA}$ at $k\gsim 1.8$ GeV. 
At large $\bk^2$ we find again good agreement with the E791 results.

\subsection{Comparison with the E791 data: $z$-distributions and 
the pion wave function}

To leading twist diffraction of pions into dijets uniquely allows 
to measure the pion distribution amplitude $\phi_{\pi}(z)$. As we 
have seen above, for moderately large $k$ studied experimentally 
higher twist corrections are very large. The  parameter 
$\langle \kappa^2_\pi(z)\rangle$ of higher twist correction from
the pion wave function varies with $z$ and is a model dependent one, 
our estimate (\ref{eq:4.2.3}) gives only a crude idea on its
$z$ dependence and (\ref{eq:4.4.1}), (\ref{eq:4.4.2})
must be regarded as numerical estimates within the factor two. 
Notwithstanding these uncertainties, even on the free nucleon target
one must be able to distinguish experimentally between the 
double-humped CZ \cite{CZ} and asymptotic distribution functions.  

To the contrary, the nuclear-rescattering driven higher twist 
correction is model-independent. The $x_{\Pom}$-$z$
correlation driven $z$ dependence of $G({1\over2} x_{\Pom},\bk^2)$
is very weak, it only slightly enhances the cross section around
$z={1\over 2}$. For this reason,
even if higher-twist dominated, the diffraction off heavy targets
is a good probe of the pion distribution amplitude $\phi_{\pi}(z)$.
The E791 paper does not
give the acceptance-corrected $z$-distributions. Ashery concludes 
\cite{SELEX} that within the $\sim 20\%$ experimental
error bars the observed E791 $z$-distribution is consistent with the 
Monte Carlo modeling based on the asymptotic pion distribution 
$\phi_{asym}(z)$. Because the observed $z$-distribution
(\ref{eq:4.2.2}) given by our model soft wave function is very 
close to the asymptotic one, it is perfectly consistent with the 
E791 data.

\section{Conclusions}

We developed the perturbative QCD description of diffraction dissociation
of pions into hard dijets on nucleons and nuclei. To leading twist
dijet excitation is shown to be dominated by the Pomeron splitting mechanism
and the two diffractive helicity amplitudes are shown to be proportional
to the unintegrated gluon structure function of the proton. We derived
an multiple-Pomeron splitting expansion of nuclear amplitudes which
is of antishadowing nature. To leading twist there is a remarkable 
cancelation of  
nuclear attenuation and antishadowing/broadening effects. We obtained
a model-independent estimate for next-to-leading twist corrections driven by 
nuclear rescatterings. These higher twist corrections are shown to be very large 
up to jet momenta $\bk^2 \gsim $ (5--7) GeV$^2$ and affect substantially the
$\bk^2$ and atomic mass number of the diffraction cross section. The model
dependence of extraction of the $z$ dependence of the pion distribution 
amplitude is shown to be weak for diffraction off heavy nuclei. Our
calculations based on the recent determination of unintegrated gluon 
structure function of the proton reproduce well the basic experimental
findings from the recent E791 experiment. A simple interpretation of 
the observed $k$ dependence is not possible though because in the E791 
range of $k$ the cross section is overwhelmed by higher twist effects.
\\

\noindent{\bf Acknowledgments:} We are grateful to Danny Ashery for 
helpful correspondence on the E791 data. Thanks are due to B.G.Zakharov 
for discussions during the early stages of this work. This research has
partly been supported by the grant INTAS 97-30494.

\pagebreak

\newpage

\begin{figure} [H]
\begin{center}
\epsfig{file=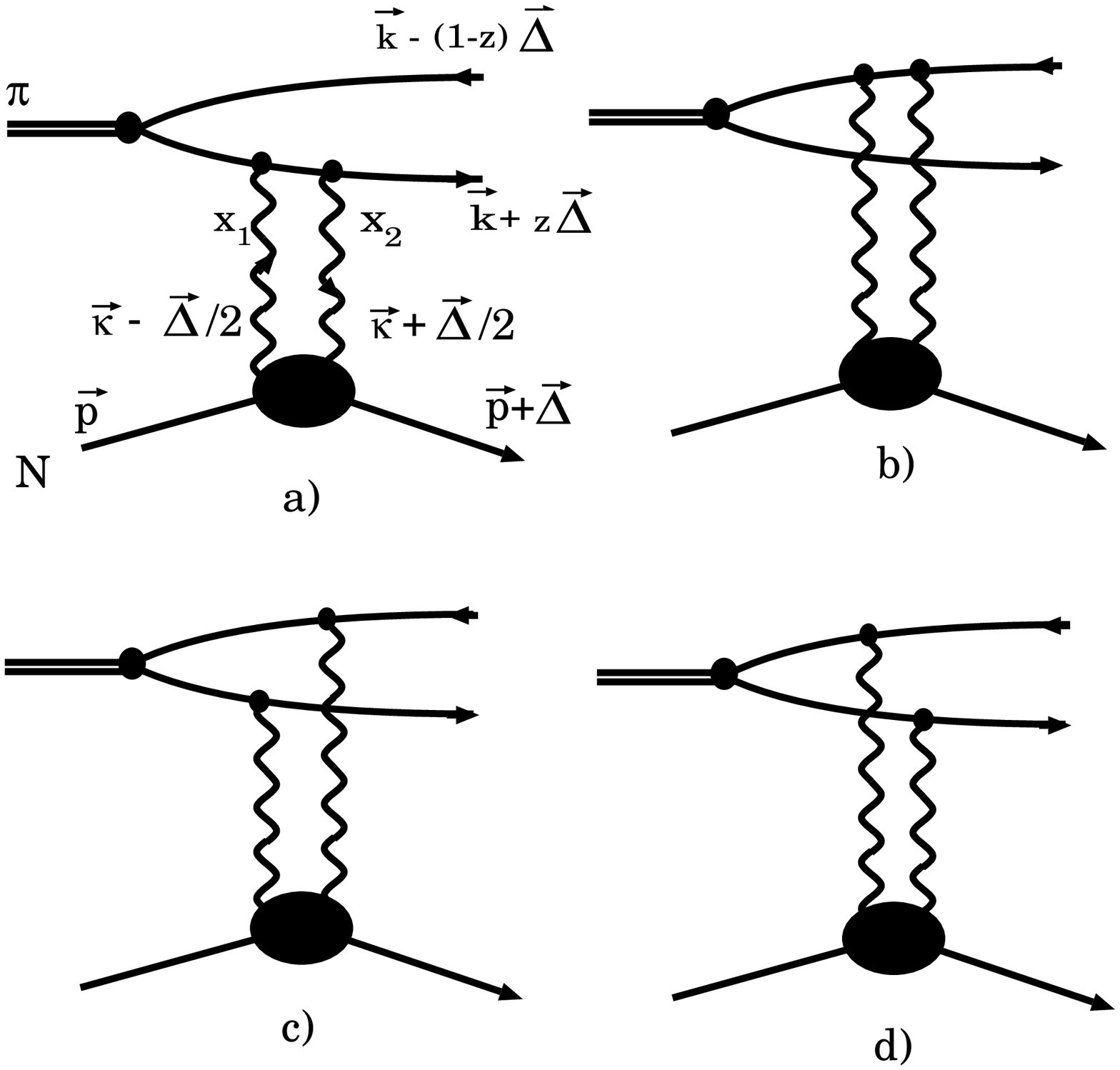, height = 14.0cm, width=14.0cm}
\end {center}
\caption{\it Feynman diagrams for diffractive dijet excitation
in $\pi$N collisions.}
\label{fig1}
\end{figure}

\newpage

\begin{figure} [H]
\begin{center}
\epsfig{file=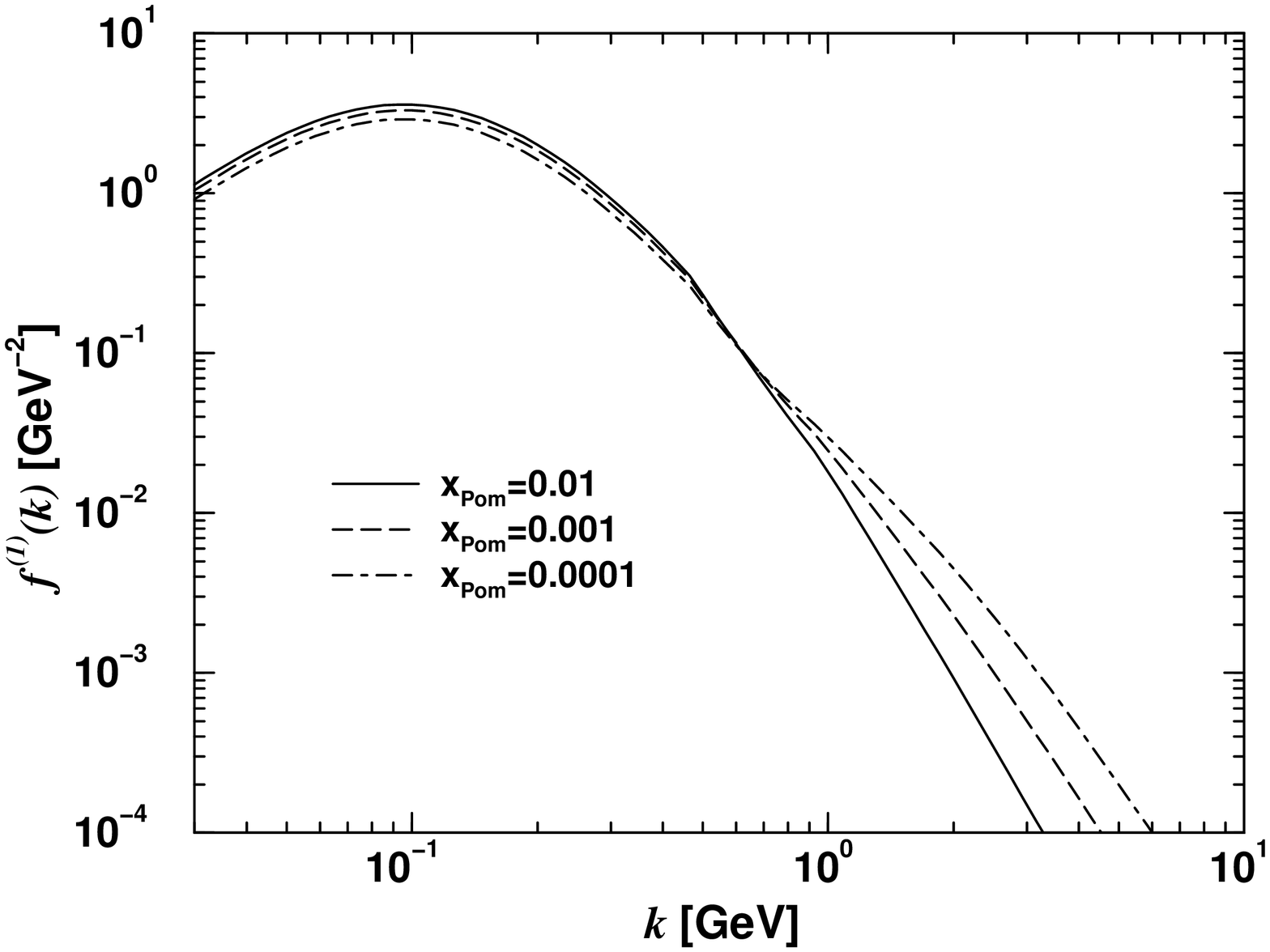, height = 14.0cm, width=14.0cm, angle=270}
\end {center}
\caption{\it The $\bk$--distribution $f^{(1)}(\bk)$ as a function of
$k$ for several values of $x_\Pom$
.}
\label{fig2}
\end{figure}

\newpage

\begin{figure} [H]
\begin{center}
\epsfig{file=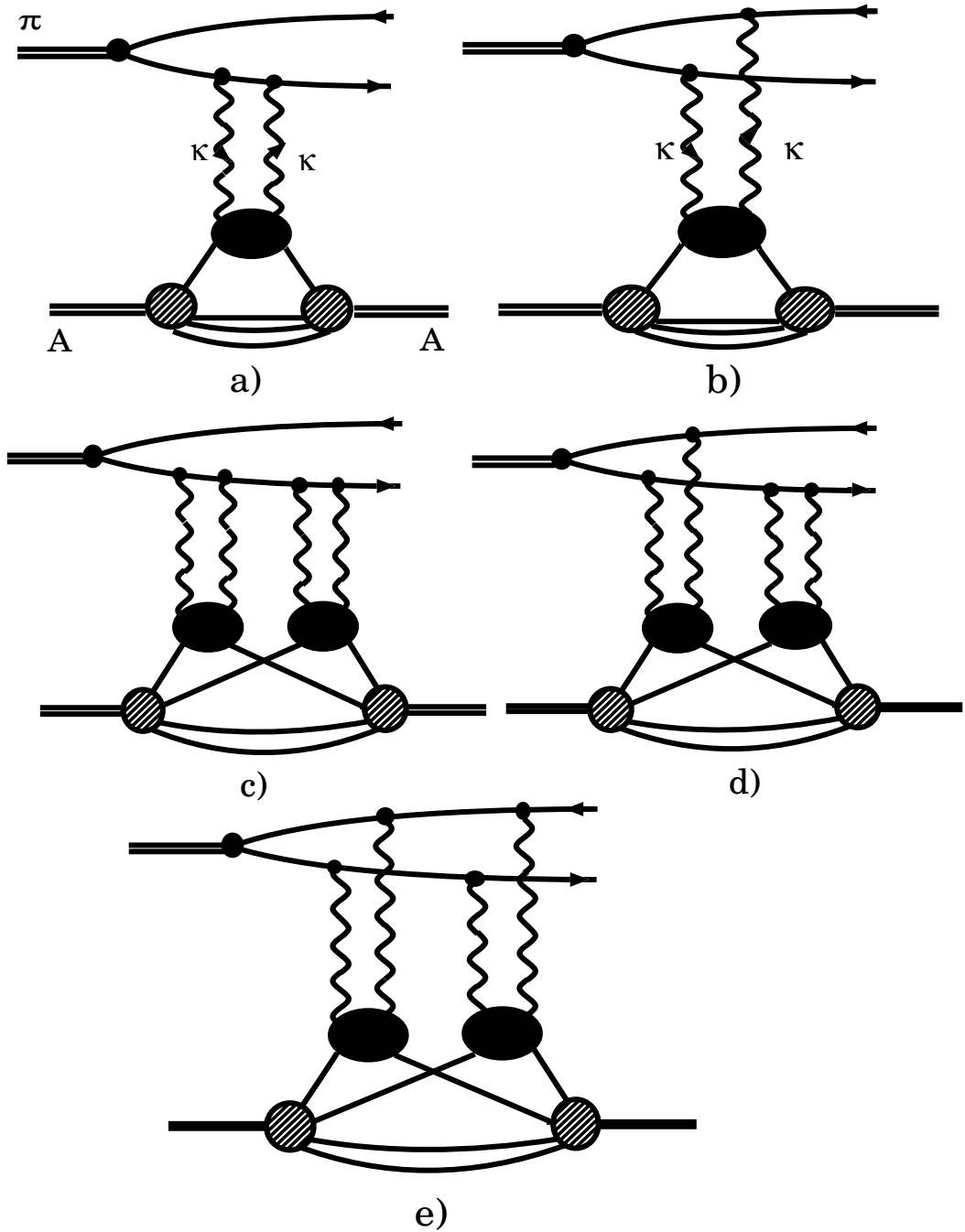, height = 18.0cm, width=14.0cm}
\end {center}
\caption{\it The nuclear multiple scattering series for 
diffractive dijet excitation on nuclei. Diagrams
 a),b)  are sample diagrams of impulse approximation,
diagrams c)--e) represent the various types of double scattering
contributions. Higher order contributions, that appear in the 
calculation are not shown.}
\label{fig3}
\end{figure}

\newpage

\begin{figure} [H]
\begin{center}
\epsfig{file=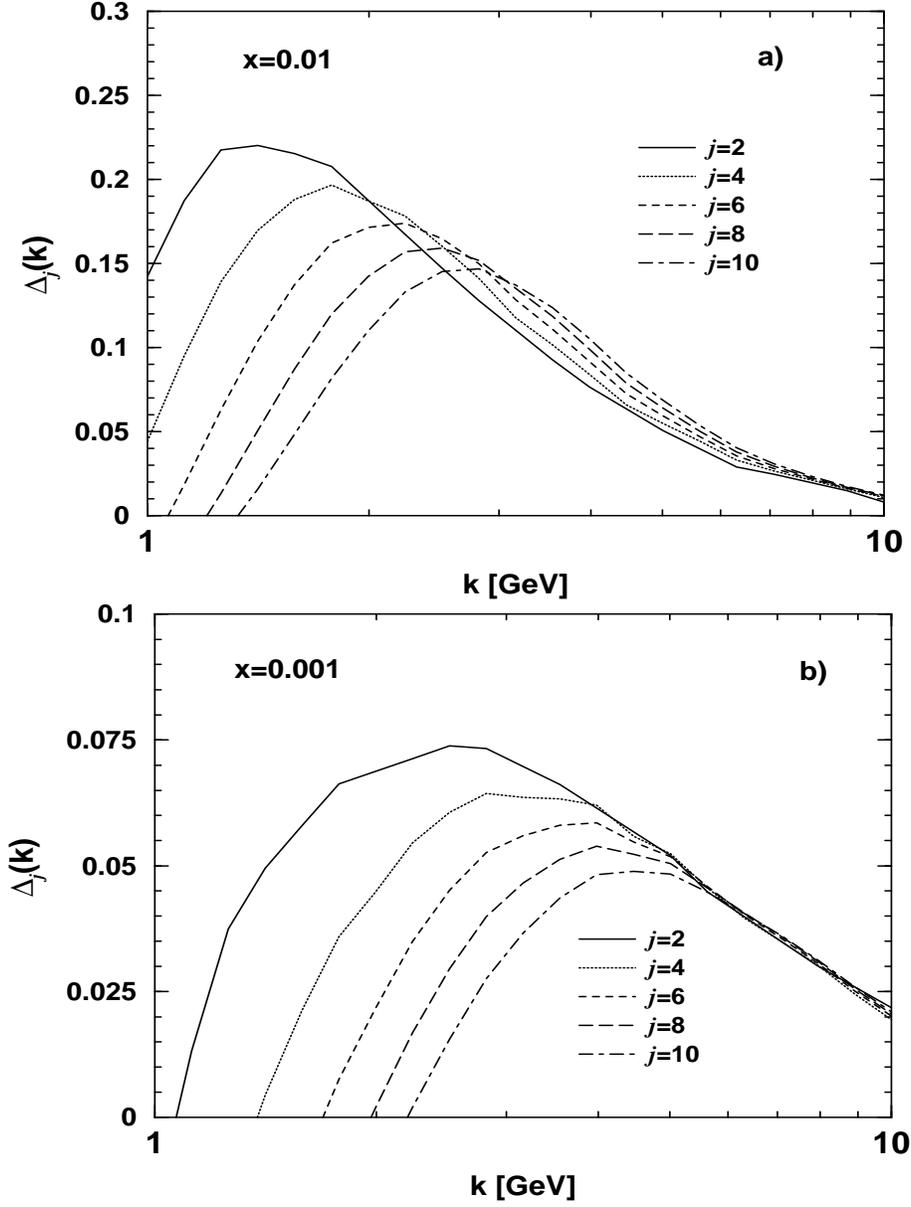, height = 16.0cm, width=12.0cm}
\end {center}
\caption{\it Large $k$--scaling properties of the multiple
convolution integrals. Shown is  $\Delta_j (k)$ for $x=0.01$
[in panel a)], and for $x=0.001$ [panel b)].}
\label{fig4}
\end{figure}

\newpage

\begin{figure} [H]
\begin{center}
\epsfig{file=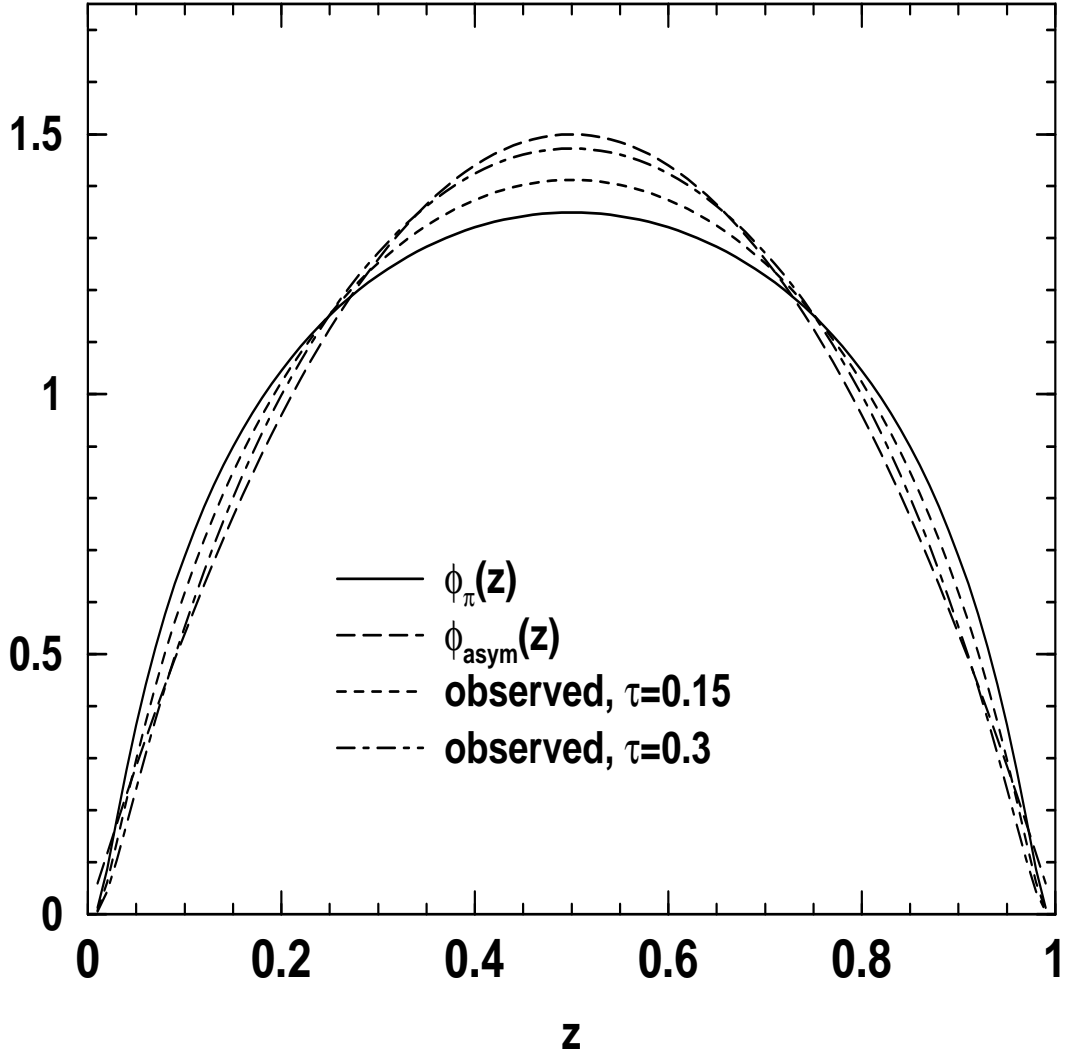, height = 14.0cm, width=14.0cm,angle=270}
\end {center}
\caption{\it The pion distribution amplitude $\phi_\pi (z)$.
The solid line is $\phi_\pi(z)$ calculated from the soft wave function 
eq.(\protect\ref{eq:4.2.1}), the dashed line is the asymptotic distribution amplitude.
The curves labeled 'observed' show the $z$--dependence of 
the soft pion distribution amplitude modulated with 
the $z$--dependence due to the kinematical $x_\Pom-z$ correlation
for two different values of the effective exponent $\tau$,
see eq.(\protect\ref{eq:4.2.4})}
\label{fig5}
\end{figure}

\newpage

\begin{figure} [H]
\begin{center}
\epsfig{file=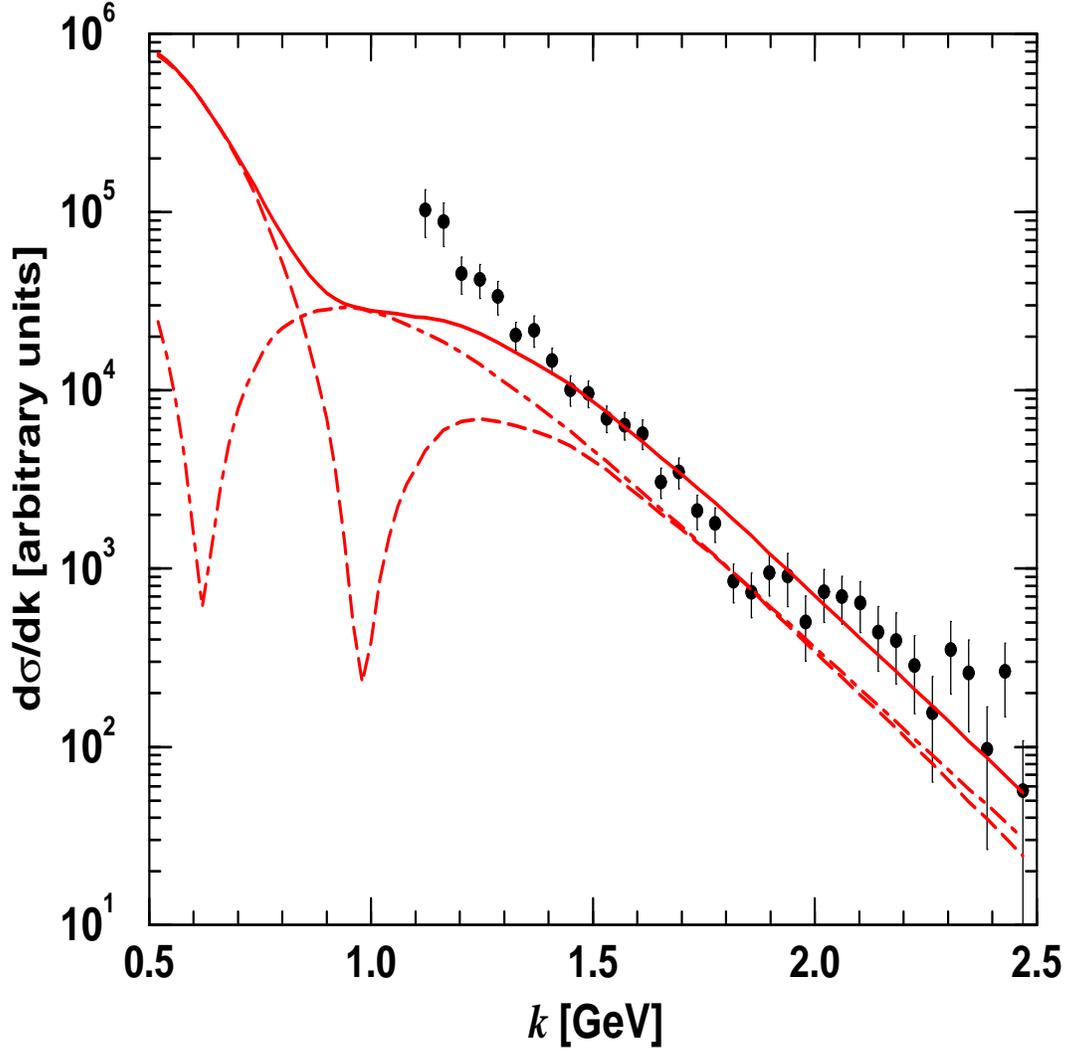, height = 14.0cm, width=14.0cm,angle=270}
\end {center}
\caption{\it The E791--data \protect\cite{SELEX} for the differential diffractive
dijet cross section $d\sigma/dk$ for the $^{196}$Pt target with
the theoretical calculations. The data are not normalized.
The dash--dotted line shows the contribution of the helicity
amplitude $\Phi^{(A)}_0$, the dashed line is
the contribution from $\bPhi^{(A)}_1$. The solid line is the 
total result.}
\label{fig6}
\end{figure}

\newpage

\begin{figure} [H]
\begin{center}
\epsfig{file=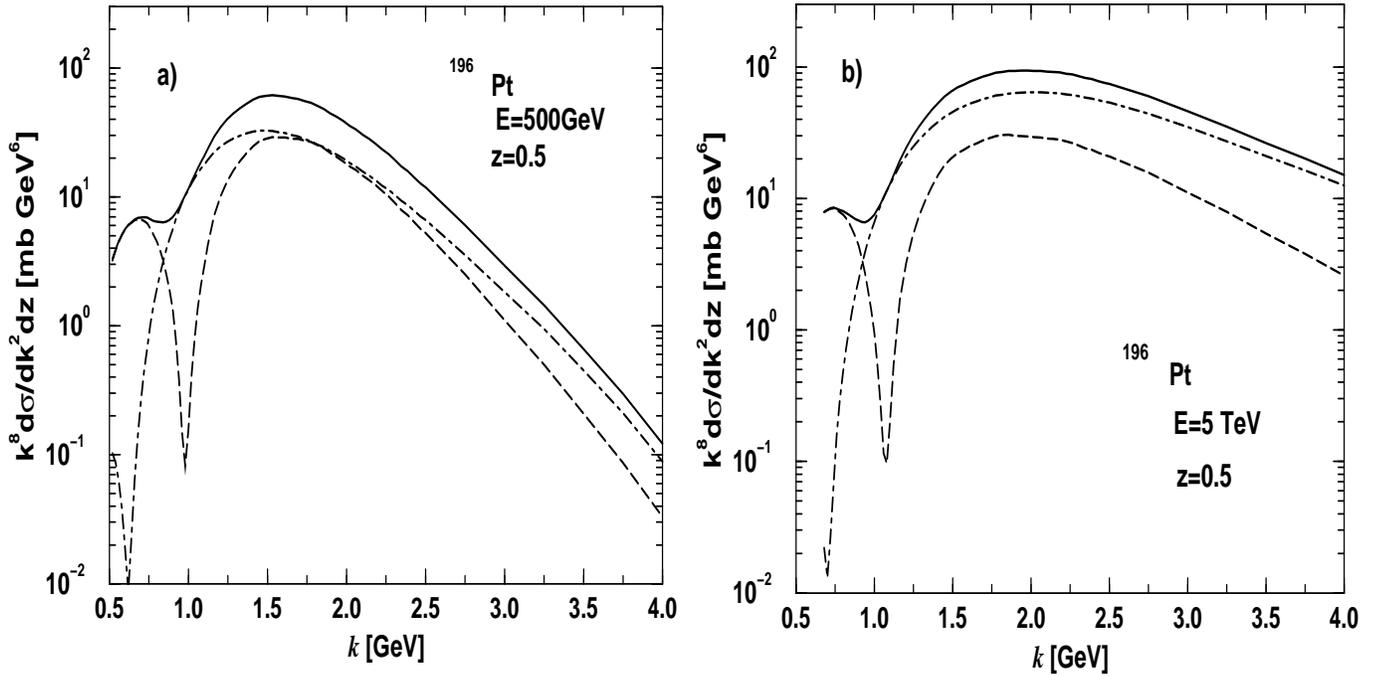, height = 18.0cm, width=9.0cm,angle=270}
\end {center}
\caption{\it Theoretical predictions for the differential dijet--cross section
for the $^{196}$Pt target.
Panel a) is for the energy of the E791 experiment, $E=500 \, GeV$, panel b)
for $E = 5 \, TeV$.The dash--dotted line shows the contribution of the helicity
amplitude $\Phi^{(A)}_0$, the dashed line is
the contribution from $\bPhi^{(A)}_1$. The solid line is the 
total result.}
\label{fig7}
\end{figure}

\newpage

\begin{figure} [H]
\begin{center}
\epsfig{file=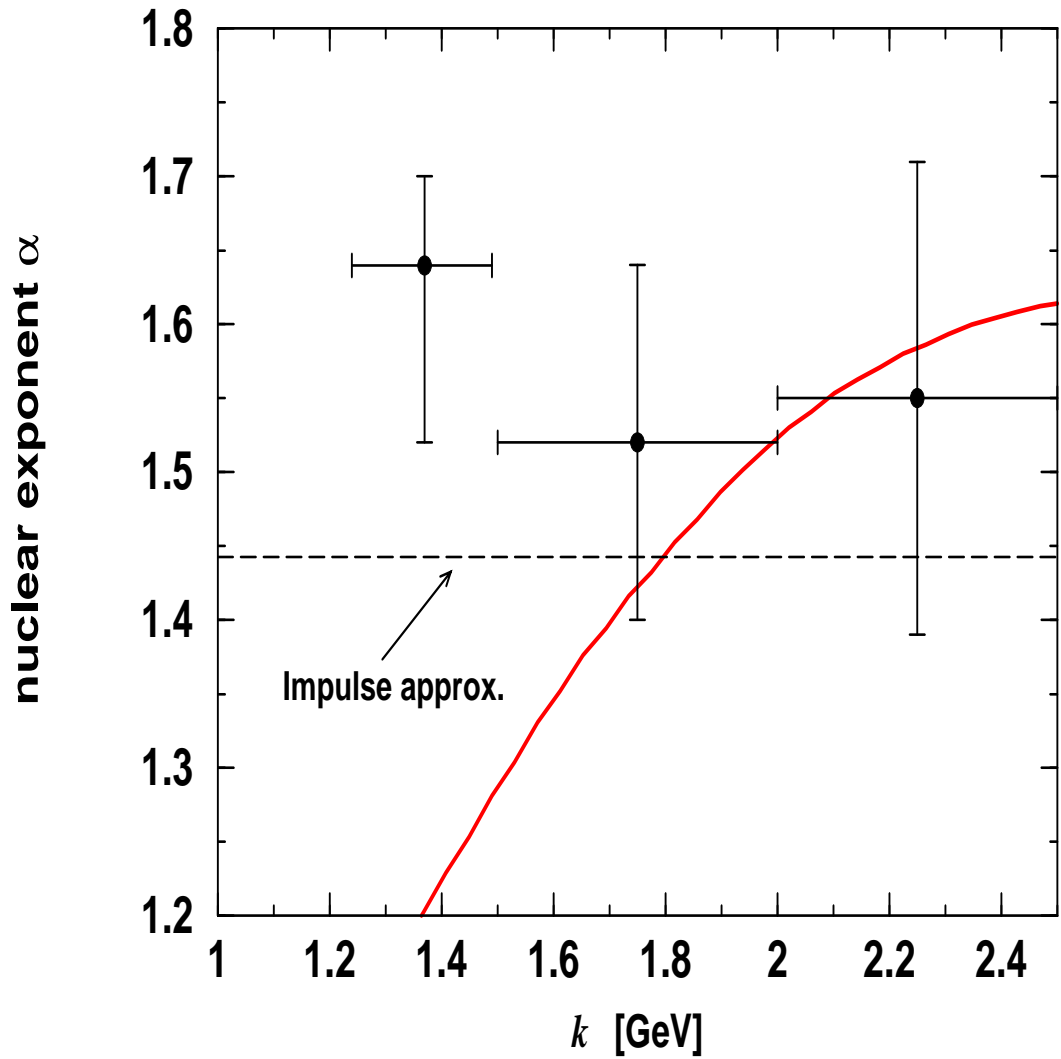, height = 14.0cm, width=14.0cm,angle=270}
\end {center}
\caption{\it Exponent $\alpha$ of the atomic mass number dependence 
of the dijet cross section with the results from E791 \protect\cite{SELEX}. 
The dashed line shows the impulse approximation 
result, the solid line is the result of the full multiple scattering series.}
\label{fig8}
\end{figure}

\end{document}